\documentclass[]{aa}

\usepackage{xcolor}
\usepackage{graphicx}
\usepackage{nicefrac}
\usepackage{float,ulem}
\usepackage[per-mode=reciprocal, multi-part-units = single, range-units = single]{siunitx}	
\DeclareSIUnit[number-unit-product = {\,}]
 \Maxwell{Mx}
\DeclareSIUnit[number-unit-product = {\,}]
 \Gauss{Gauss}
\DeclareSIUnit[number-unit-product = {\,}]
 \pixel{pixel}
\usepackage{natbib}
\bibpunct{(}{)}{;}{a}{}{,} 
\usepackage{txfonts}
\usepackage[colorlinks]{hyperref}
\hypersetup{
colorlinks=true,
    citecolor={blue},
    linkcolor={blue}
}
\usepackage{cleveref}
\crefformat{footnote}{#2\footnotemark[#1]#3}

\begin{document} 

 \title{Evolution of solar surface inflows around emerging active regions}

 \author{N. Gottschling\inst{1}
          \and
          H. Schunker\inst{1}\fnmsep\inst{2}
          \and
          A. C. Birch\inst{1}
          \and
          B. L\"optien\inst{1}
          \and
          L. Gizon\inst{1}\fnmsep\inst{3}\fnmsep\inst{4}
          }

 \institute{Max-Planck-Institut f\"ur Sonnensystemforschung,                           Justus-von-Liebig-Weg 3, 37077 G\"ottingen, Germany\\
            \email{gottschling@mps.mpg.de}
        \and 
            School of Mathematical and Physical Sciences, University of Newcastle, Callaghan, New South Wales, Australia
            \and
            Institut f\"ur Astrophysik, Georg-August Universit\"at G\"ottingen, 37077 G\"ottingen, Germany
            \and 
            Center for Space Science, NYUAD Institute, New York University Abu Dhabi, Abu Dhabi, UAE
            }

   \date{}

  \abstract
   {Solar active regions are associated with Evershed outflows in sunspot penumbrae, moat outflows surrounding sunspots, and extended inflows surrounding active regions. The latter have been identified on established active regions by various methods. The evolution of these inflows and their dependence on active region properties as well as their impact on the global magnetic field are not yet understood.}
   {We aim to understand the evolution of the average inflows around emerging active regions and to derive an empirical model for these inflows. We expect that this can be used to better understand how the inflows act on the diffusion of the magnetic field in active regions.}
   {We analyze horizontal flows at the surface of the Sun using local correlation tracking of solar granules observed in continuum images of the Helioseismic and Magnetic Imager (HMI) onboard the Solar Dynamics Observatory (SDO). We measure average flows of a sample of 182~isolated active regions up to seven days before and after their emergence onto the solar surface with a cadence of 12~hours. About half of the active regions in the sample develop sunspots with moat flows in addition to the surrounding inflows. We investigate the average inflow properties with respect to active region characteristics of total flux and latitude. We fit a model to these observed inflows for a quantitative analysis.}
   {We find that converging flows of around \SIrange{20}{30}{\meter \per \second} are first visible one~day prior to emergence, in agreement with recent results.
   These converging flows are present independently of active region properties of latitude or flux.
   We confirm a recently found prograde flow of about \SI{40}{\meter \per \second} at the leading polarity during emergence.
   We find that the time after emergence when the latitudinal inflows increase in amplitude depends on the flux of the active region, ranging from one to four days after emergence and increasing with flux.
   The largest extent of the inflows is up to about \SI{7\pm1}{\degree} away from the center of the active region within the first six days after emergence. The inflow velocities have amplitudes of about \SI{50}{\meter \per \second}.}
{}


\maketitle


\section{Introduction}
\label{Sec_Introduction}

Active regions (hereafter ARs) are patches of magnetic field at the surface of the Sun. They are the host to sunspots visible in white light images \citep{Driel_Gesztelyi_2015} and the site of a large variety of phenomena, such as flares and coronal mass ejections. Theories of where and how active regions form include a buoyant rise of magnetic flux tubes from the bottom of the convection zone (see review by \citealt{Fan_LRSP_2009}), formation in the bulk of the convection zone \citep{Nelson_2013}, or in the near-surface layers \citep{Brandenburg_2005}.

At the surface, active regions are advected on large spatial scales by differential rotation \citep{Snodgrass_1983} and the meridional flow \citep{Duvall_1979}. On smaller scales, magnetic field is diffused by the turbulent convective motions of granulation and supergranulation \citep{Leighton_1964}. Methods to infer the large-scale flows include tracking the motions of small-scale convective or magnetic features, which are displaced by the larger-scale motions, as well as a variety of tools for the analysis of solar oscillations, which are known as local helioseismology (see review by \citealt{Gizon_LRSP_2005}).

In addition, \citet{Gizon_2001} found that evolved active regions are surrounded by horizontal converging flows, using time-distance helioseismology \citep{Duvall_1993} to study flows near the surface of the Sun. These surface inflows typically extend to about \SI{10}{\degree} from the active region and have horizontal velocities of up to \SI{50}{\meter \per \second} \citep{Gizon_2001, Haber_Hindman_2004, Zhao_2004, Hindman_2009, Loeptien_2018, Braun_2019}. What drives these inflows is an interesting question.
One possible mechanism is that enhanced emission of radiation in the magnetic field structures leads to a horizontal temperature gradient which could drive the flows \citep{Spruit_2003}. 

The inflows towards active regions are important as they affect the advection and flux cancellation of ARs, and may counteract diffusion \citep{DeRosa_Schrijver_2006}. \citet{DeRosa_Schrijver_2006} and \citet{MartinBelda_2017_Evolution, MartinBelda_2017_Inflows} included models of the inflows into surface flux transport simulations, showing that they can provide a feedback mechanism which saturates the amplitude of the global magnetic field. This could explain observed variations in solar cycle strengths. 

Several studies applied local helioseismic techniques to investigate different aspects of the properties of these flows. \citet{Haber_Hindman_2004} studied their depth dependence, identifying a transition to outflows in deeper layers for some AR. \citet{Komm_2011,Komm_2012} studied vertical flows and vorticity, finding that emerging flux is associated with weak upflows of less than \SI{1}{\meter \per \second} and increasing strength of vorticity, and that decaying flux is associated with downflows and decreasing strength of vorticity.

Recently, \citet{Braun_2019}, using helioseismic holography \citep{Lindsey_2000}, investigated a sample of 336~ARs and derived flow patterns for subgroups of increasing magnetic flux.
This work showed that the amplitude of the flows increases with magnetic flux, and reported a retrograde flow component mainly at the poleward side and, less distinct, the equatorward side of the AR.

\citet{Loeptien_2017} used Local Correlation Tracking (LCT, \citealt{November_Simon_1988}) of the convective granulation pattern on the solar surface to measure inflows around an average high-flux AR. The regions in their sample have fluxes larger than \SI{5.9e21}{\Maxwell}. Their work confirms inflows of about \SIrange{20}{30}{\meter \per \second} spanning about \SI{10}{\degree} from the AR.

\citet{Birch_2019} found a compact (less than \SI{2}{\degree}) converging flow of about \SI{40}{\meter \per \second} in the day before active region emergence, using both LCT and helioseismic holography as independent methods for flow inferences on the surface. Comparisons of flows derived with LCT and local helioseismology showed that they are in good agreement \citep{Svanda_2013, Birch_2016}.

In this study, we aim to measure the evolution and structure of the inflows around active regions at the solar surface during the early stages of the active-region evolution, that is before, throughout, and in the days after emergence. We also investigate potential differences in the inflows related to the total unsigned flux and the latitude of the active regions.

We perform a statistical study by averaging over a large sample of ARs. For this, we make use of the SDO Helioseismic emerging active regions (SDO/HEAR) survey \citep{Schunker_2016}. By using this sample rather than performing a case study, we decrease background noise as well as peculiarities of individual emerging active regions (EARs), which can have a dominating effect on the flow structure.

The paper is structured as follows: In Sect.~\ref{Sec_Data} we describe the sample of emerging active regions as well as the flow measurements. Sect.~\ref{Sec_Processing} illustrates the data processing that we applied to derive the evolution of flows around ensemble averages of ARs. Sect.~\ref{Sec_Results} presents our results, followed by a discussion (Sect.~\ref{Sec_Discussion}).

\section{Data}
\label{Sec_Data}

\subsection{The sample of emerging active regions}

\citet{Schunker_2016} describe the HEAR survey in detail. Here, we summarize the main aspects.

The HEAR survey comprises 182~emerging active regions during the time from 2010 to 2014. They were selected from the National Oceanic and Atmospheric Administration (NOAA) solar region reports on the basis of several criteria, requiring that the AR
a)~reaches a total sunspot area of at least 10~micro~hemispheres ($\mu$H; 1\;$\mu$H $\approx$ \SI{3}{\square \mega \meter}), b)~has its first NOAA record within \SI{50}{\degree} of the central meridian, and
c)~emerges into the quiet Sun, without pre-existing large-scale flux within a radius of \SI{18}{\degree}.
These criteria were chosen to minimize contamination by pre-existing magnetic field, and to reduce projection effects. The emergence time of each active region is defined as the time at which the region has reached \SI{10}{\%} of the maximum unsigned flux within the first \SI{36}{\hour} after the active region is first recorded by NOAA. The survey also provides the Carrington longitude and latitude as well as the start and end times of the observations for each region.

In addition to the emerging active regions, the survey includes one quiet Sun control region for each EAR. These control regions match their corresponding EARs in latitude and distance from the central meridian at a mock-emergence time, at which there are no numbered SHARP regions within a radius of \SI{18}{\degree} of the control region. These control regions are important as a comparison against which any measurement on the EARs can be tested.

\subsection{Flows inferred from local correlation tracking}
\label{subsect_Flows_from_LCT}

We use an updated version of the flow maps generated by \citet{Loeptien_2017}, who describe their processing procedure in detail. Here, we summarize the important aspects.

\citet{Loeptien_2017} applied Local Correlation Tracking on full-disk continuum intensity images from SDO/HMI \citep{Schou_2012}. Specifically, they ran the FLCT code \citep{Welsch_2004, Fisher_Welsch_2008} on hmi.ic\_45s series data, starting on 24~April~2010 and ending on 27~April~2016. In order to reduce computation time, only one flow measurement every 30~minutes was calculated, by cross-correlating pairs of images that are 45~seconds apart. Each measurement consists of a map of velocities $v_x$ in $x$-direction and a map of velocities $v_y$ in $y$-direction (with $x$ and $y$ in CCD coordinates).

LCT is known to underestimate flow velocities (see e.g. \citealt{Loeptien_2016_DataCompression, Verma_2013, Svanda_2007} and references therein). This was addressed by generating calibration data with which the flow velocities were then corrected (see Sect.~2.1 of \citealt{Loeptien_2017}).

The noise in LCT calculations increases to the limb, due to projection effects. Therefore, the flow maps are cropped at \SI{60}{\degree} from disk center.
The maps suffer from systematic effects, such as the orbital motion of HMI and the shrinking-Sun effect \citep{Lisle_2004, Loeptien_2016_shrinkingSun}, a systematic converging flow towards disk center.
This was addressed by representing the background signal of the flow maps as a sum of low-order Zernike polynomials (up to radial degree $n=7$). The time series of the Zernike polynomial coefficients were then transformed into Fourier space, where frequencies corresponding to known periods of systematic effects were isolated. These are the 24~hour period of the satellite orbit and its 23~observed harmonics, and the 365~day period of the earth orbit and its first 16~harmonics (down to a period of 22.8~days), which show significant power. The filtered, backtransformed background estimate was then subtracted from the observations.

An exception was made for the time series of Zernike polynomials $Z_1^{-1} = y$ for the $v_x$ maps and $Z_1^{1}= x$ for the $v_y$ maps, which are correlated to each other and were therefore not treated in the way outlined above. For these, the first element of a principle component analysis of the two components was subtracted (as described in Appendix~B of \citealt{Loeptien_2017}). However, this retained a systematic variation for these terms, on the order of several~m~s${^{-1}}$ with a period of 24~hours and its harmonics, which corresponds to large-scale gradients in the flow maps. To correct for this, for the present study we repeated the data processing of \citet{Loeptien_2017}, subtracting for the Zernike polynomials $Z_1^{-1} = y$ for $v_x$ and $Z_1^{1}= x$ for $v_y$ the same harmonics as for the other components. In this rerun, the data coverage was increased to 6~March~2020.

As a last step, we remapped the velocity maps from the CCD coordinate system to Plate Carree projection (Carrington longitude, latitude), with a total size of 3600~x~1200 grid points and a grid spacing of \SI[number-unit-product =]{0.1}{\degree}. In this projection, averaging over regions at different latitudes and longitudes can be done in a straightforward manner. In the same step, the velocities were transformed from $v_x, v_y$ to the longitudinal and latitudinal velocities $v_{\rm lon}, v_{\rm lat}$. We then applied a Gaussian filter with a width of \SI{0.4}{\degree} and subsampled to the grid spacing of \SI{0.4}{\degree} used by \citet{Loeptien_2017}. This processing step differs from \citet{Loeptien_2017}, in order to remove aliasing from the data.

\section{Data reduction}
\label{Sec_Processing}

\subsection{Processing and validation of the flow data}
\label{subsec_processingvalidation}

In the flow maps, individual pixels occasionally show unrealistically high values. This affects about 10~pixels per map, mostly close to the limb. We speculate that these arise from the decreased contrast at the limb in combination with the evolution of the granules. To mitigate their effect, we replace outlier pixels that are outside five standard deviations of the mean of the velocity distribution by the mean value of pixels surrounding them in a box of \SI{5x5}{\pixel}.

In addition to the systematic effects described above (the orbital motion of SDO and the shrinking-Sun effect), the LCT data exhibits another long-term, large-scale modulation: At the beginning of the time series in 2010, the longitudinal flow component $v_{\rm lon}$ has a bias towards prograde (retrograde) velocities in the eastern (western) half of the disk, and the latitudinal flow component $v_{\rm lat}$ has a bias towards southward (northward) velocities in the northern (southern) half of the disk, mostly at latitudes above \SI{30}{\degree}. Over the years, this modulation changes, such that the hemispheric biases of the two flow components have reversed sign around the year 2015. The amplitude of this modulation is about \SI{10}{\meter \per \second} in a half-year average. The reason for this modulation is not fully understood. Its time scale does not correspond to any known periodic effects, such as the B angle or the orbital motion of HMI.

To correct for this large-scale modulation, we translate all velocity maps from Carrington longitude to Stonyhurst longitude, create running time averages over half a year of data each, centered ten days apart, and subtract these from every individual map from five days before until five days after the central time of the average. The corrected maps are then mapped back to Carrington longitude. We choose an averaging time of half a year because a shorter time scale would potentially subtract real flows, and a spacing of ten days to avoid drastic changes between two blocks of subtraction. The average maps are smoothed with a Gaussian of $\sigma= \SI{0.8}{\degree}$ to reduce small-scale effects.

With the above processing, which was necessary to address the various systematic errors (the orbital motion of HMI, the shrinking-Sun effect, the long-term modulation), the velocities that we measure throughout our analysis are therefore relative to velocities $v_{\rm lon}=0$, $v_{\rm lat}=0$, which are the half-year time averages at each location in Stonyhurst longitude and latitude.

As a test of the above processing, we want to make sure that the Zernike subtraction outlined by \citet{Loeptien_2017} does not subract the kind of flows we want to study together with the systematic effects. To evaluate how much this procedure alters the inflows, we create one Carrington rotation of synthetic flow data with inflow features and apply the same filtering on them as was done on the actual data. Appendix~\ref{section_Appendix_synthetic_data} describes this test. We find that the extent and amplitude of the synthetic flows is well preserved, with changes of less than \SI{5}{\meter \per \second} for model inflows with an amplitude of \SI{50}{\meter \per \second}. Small deviations are introduced to the data, with amplitudes of less than \SI{5}{\meter \per \second}.

As another test of the LCT flows used in this study, we carry out a control against an independent data set. In this test, we project the LCT flows to the line-of-sight component and compare them to flow measurements obtained from direct Doppler images from SDO/HMI. Appendix~\ref{section_Appendix_vLOS_vDopp} describes this test. The Dopplergrams only provide one velocity component, and therefore are not suited as the main flow measurement of this study. We find that in general, the two methods agree well, with correlations exceeding 0.8, except for in close proximity (\SI{2}{\degree}) to sunspots. This might be due to the different depths at which the two measurements are taken, as well as different sensitivities in the presence of strong magnetic field.

To use the LCT flow data on the sample of emerging active regions and control regions, for each region we extract data cubes of the two flow components $v_{\rm lon}$, $v_{\rm lat}$, which are centered at the position of the region in Carrington longitude and latitude given in the HEAR survey, and span $\pm$\SI{30}{\degree} in both longitude and latitude. The cubes start (end) at the full or half hour before (after) the first (last) observation time given in the HEAR survey, respectively. This is because the HEAR survey defined the emergence time from the HARP vector magnetogram observations that have a cadence of 12~minutes, whereas the LCT data has a cadence of 30~minutes, with data at full and half hours. For the same reason, in this work we define the time of emergence as the first full or half~hour after the time of emergence $t_0$ in the HEAR survey. We chose this to make sure that the emergence has occurred for all regions consistently.

The typical observation period of an individual region is about nine~days (450 frames). The maximum difference between the start (end) of the observations and the time of emergence is seven~days (343~frames) before (after) emergence (leaving about two~days of observations after (before) emergence). The averaging over ensembles of active regions is done relative to the time of emergence (see Sect.~\ref{Sec_Results}). The total duration of observations therefore is 14~days, with a varying number of active regions at each time relative to the time of emergence. The parts of each frame that are off the visible disk are not included in any averaging.

For the further analysis of the characteristics of the EARs, for each region we generate magnetogram and continuum intensity data cubes that match the setup of the flow data described above. We use the same observation start and end times, with the same 30~minute cadence as for the flow maps (described above) on the continuum intensity series hmi.ic\_45s to create the intensity data cubes.
The data is remapped to Plate Carree projection, with a grid spacing of \SI[number-unit-product =]{0.1}{\degree}. This is four times higher than that of the flow maps, and chosen for both continuum and magnetogram data to preserve small-scale information. We correct for limb darkening and normalize the continuum data cubes by convolving each frame with a Gaussian of $\sigma=\SI{5}{\degree}$, to retrieve the large-scale variation in the frame (the limb darkening profile), and then dividing the frame by this background. For the magnetograms, we extract the full-vector magnetic field from hmi.b\_720s. From the vector field components, we calculate the radial field $B_z$, using the transformation given by \citet{Gary_1990}. For pixels where the field azimuth is not calculated with the minimum-energy algorithm, we use the random field disambiguation \citep{Hoeksema_2014}. We then average together the 5~frames of the 720~s radial field $B_z$ data within one hour.

The data of each EAR thus consists of three-dimensional data cubes (time, longitude, latitude) for the flow components in longitudinal and latitudinal direction, the radial magnetic field, and the continuum intensity. Each cube has frame sizes of \SI{60x60}\degree in longitude and latitude, centered at the Carrington longitude and latitude of the AR given in the HEAR survey, and covers about 14~days, centered on the time of emergence, with data coverage of the visible disk for around nine~days. The cubes of the velocity maps and the continuum intensity have a cadence of 30~minutes, the cubes of the radial magnetic field have a cadence of 1~hour.

\subsection{Measuring the location of the magnetic polarities}
\label{Sec_Positions}

Over time, the regions move relative to the position given in the HEAR survey. To perform a sensible ensemble average of the active regions relative to their center or their leading or trailing polarity, we measure the positions of the leading and trailing polarities in longitude and latitude in each one-hour radial magnetic field average.
To reduce sensitivity to small-scale structure within the AR, for this measurement we smooth the magnetograms with a Gaussian of $\sigma = \SI{0.4}{\degree}$. 

As an initial guess, for each of the 182~AR we estimate the position of the polarities at one day after emergence by eye, at which time all regions show a clear bipolar structure. The procedure for each time frame in each of the 182~data cubes is then as follows:
\begin{itemize}
\item Estimate the fluctuation in the radial magnetic field as the standard deviation in a box of \SI{12x12}{\degree} around the position in Carrington longitude and latitude from the HEAR survey. The magnetic field of the AR is part of the estimate. The size of the box is chosen to exclude other AR from the field of view which would bias the estimate.

\item Consider the negative and positive polarities separately. Depending on the AR, the positive (negative) field corresponds to the leading (trailing) polarity or vice versa.

\item Identify regions as neighboring positive (negative) valued grid points (for brevity called pixels in this context) above (below) a threshold value, taken as $1.5$~times the standard deviation. We use this adaptive threshold because the levels of magnetic activity in the vicinity of the individual ARs differ. The factor of $1.5$ is applied to further reduce detection noise and is a result of trial and error.

\item Exclude regions which are only one pixel in size. These are considered to be too small for a reliable polarity position estimate.

\item Exclude regions whose flux-weighted center is further away from the initial (or first-guess, see below) position than \SI{\pm 2}{\degree}, to avoid selecting unrelated surrounding magnetic field. This is important especially during emergence, where the field of the AR is still weak.

\item Of the remaining regions: Identify the one which has the largest absolute sum in $B_z$. Define the position of the active region polarity as the longitude and latitude of its flux-weighted center.
\end{itemize}

This method is iterated forward and backward in time, starting from the frame one day after emergence, where we identified a first-guess position by eye. For each frame the position from the previous frame is used as first guess.

With this procedure, we determine the positions in longitude and latitude of each polarity for each EAR at each time step. Before emergence, by definition there is only very little magnetic flux of the active region present, which makes a position measurement unreliable, with the risk of identifying surrounding small-scale field. For consistency, we therefore set the positions in all frames before emergence to the values at emergence. We point out that other choices could be made, for example extrapolating the post-emergence motion to the times before emergence. This would however require additional assumptions on the proportionality of the motion over a certain (flux dependent) time interval, and is therefore not attempted here.

In some cases, the inferred positions in a single frame differ largely from those before and after, for example when a connection between two regions is established by pixels which have magnetic field strengths just above the threshold in the particular frame, and below in the previous and the following frame. Such artifacts will be present for any choice of threshold. To mitigate this, we smooth the time series of positions with a Gaussian of width $\sigma=\SI{3}{hours}$. We define the position of the center as half the distance between the leading and the trailing polarity positions. The total motion of an AR over the course of the observations is typically a few degrees. The largest displacements are on the order of \SI{6}{\degree} for ARs at high latitudes, at which the difference between Carrington rotation rate and solar rotation rate at the surface is largest.

Other methods to infer the position of AR polarities are possible. To estimate the difference that this might make for our analysis, we apply the method of \citet{Schunker_2019}. Their algorithm iteratively calculates the field strengths in the surroundings of all grid points, and selects that patch as the polarity which lies closest to the one in the previous frame. Since in our method we select the strongest patch instead, this can lead to differences between the two methods, because in some ARs, the dominant patch of magnetic field of one polarity is replaced later on by another patch. Appendix~\ref{section_Appendix_Positions} compares the two methods.
We applied the positions of both methods to ensemble averages of the flow data (see following sections) and found a standard deviation of the difference of the methods of below \SI{2}{\meter \per \second} close to emergence, and no systematic difference with respect to the inflows we want to investigate.

\subsection{Weight maps for ensemble averages}

\begin{figure}
\centering
\includegraphics[width=\hsize]{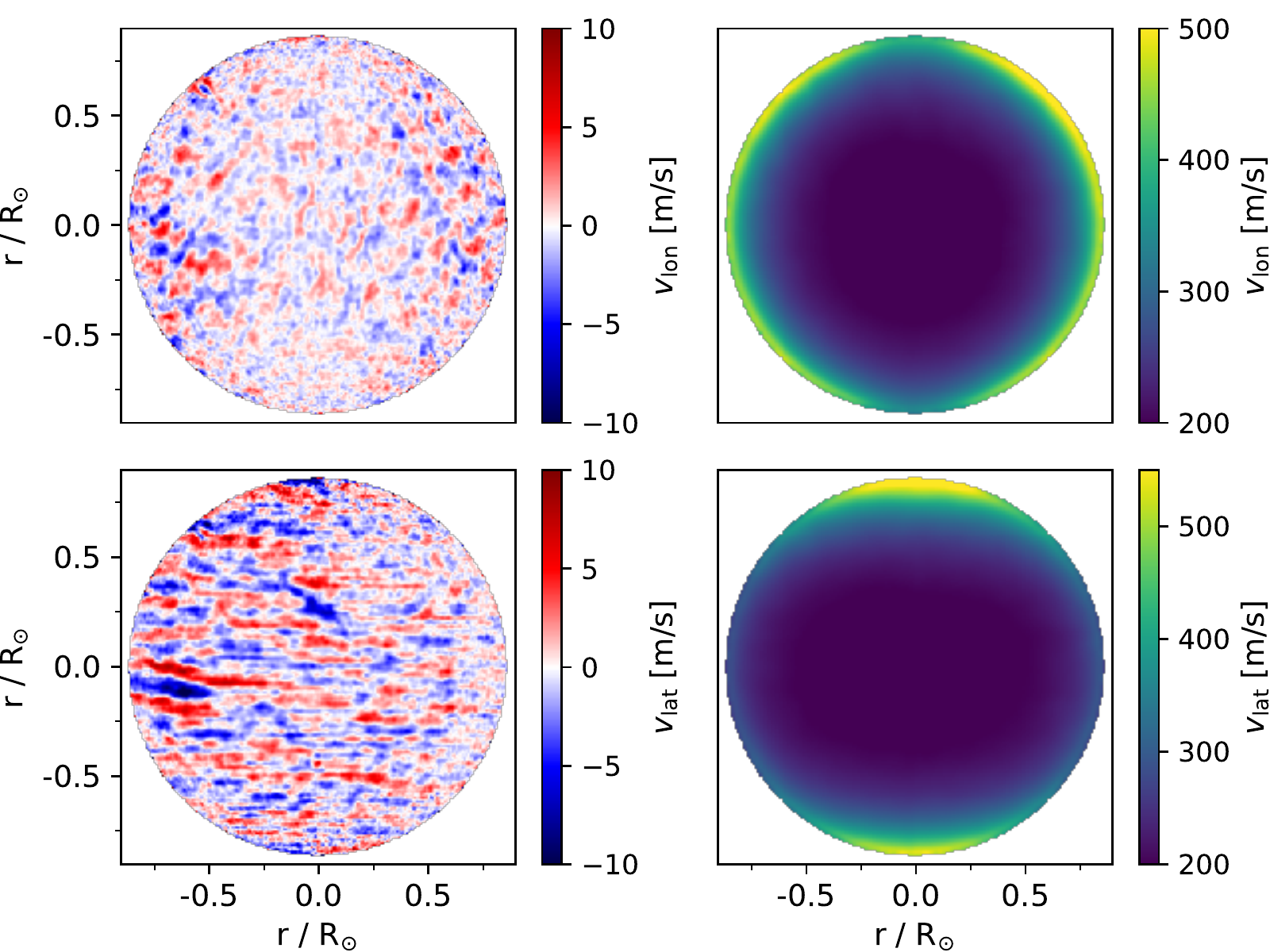}
  \caption{Temporal averages (left panels) and standard deviations (right panels) over the full-disk flow maps in the time from 24~April~2010 to 6~March~2020. The top and bottom rows show the longitudinal and latitudinal flow component, respectively.}
      \label{weightmap_stdev_average}
\end{figure}

When performing ensemble averaging, we have to take into account that at a given time relative to the time of emergence, the active regions have different disk positions and therefore different background noise levels, as the noise increases towards the limb. We account for this with weighted ensemble averages. For this, we create data cubes of weight maps for all EARs and control regions in the following way:

First, we map the full disk velocity field $v_{\rm lon}$, $v_{\rm lat}$ from the Plate Carree grid with 900~x~300 grid points and a grid spacing of \SI{0.4}{\degree} (see Sect.~\ref{subsect_Flows_from_LCT}), from Carrington coordinates into heliocentric Cartesian coordinates \citep{Thompson_2006}, centered on the Carrington longitude and latitude at disk center. Transforming into this system accounts for the B~angle and the P~angle.
We map with a linear interpolation onto a fixed grid with 300~x~300 grid points and a grid spacing of 0.006~solar radii ($R_\odot$)(corresponding to about \SI{4.2}{\mega\meter} at disk center). The frame size is \SI{1.8}{R_\odot}, which is slightly larger than the LCT data coverage (the \SI{60}{\degree} from disk center of the LCT data corresponds to $2 \sin{\nicefrac{\pi}{3}}= \SI{1.73}{R_\odot}$).

We did this for the whole ten years of LCT data. We calculate the average flows $\langle v_{\rm lon} \rangle$ and $\langle v_{\rm lat} \rangle$, and a map of the standard deviation of each flow component (see Fig.~\ref{weightmap_stdev_average}) on the fixed grid in heliocentric Cartesian coordinates. The velocities in the average maps are on the order of \SI{10}{\meter \per \second}, which is much smaller than the velocities in an individual map (on the order of \SI{1000}{\meter \per \second}). This is because the velocities are equally distributed around zero, with the dominant noise from granulation and supergranulation.

For the longitudinal component, the standard deviation increases mainly towards the east and west limbs. For the latitudinal component, the northern and southern polar regions show the largest standard deviation.

We generate full disk weight maps by mapping the standard deviation $\sigma$ of $v_{\rm lon}$ and $v_{\rm lat}$ back onto the Plate Carree projection in Carrington coordinates. From these, we create data cubes of the weight maps as $\nicefrac{1}{\sigma^2}$ for each of the 182~EARs and control regions, in the same way as the data cubes of the velocity maps (see Sect.~\ref{Sec_Processing}).

\subsection{Exclusion of magnetic pixels}
\label{subsec_blankout}

Local Correlation Tracking of solar granules is known to be unreliable within regions of strong magnetic field, where the granulation is attenuated. This is the case for sunspots or plage, which appear darker or brighter than the quiet Sun in continuum images. The motion that LCT tracks in these regions is therefore a combination of the motion of the granules and the motion of these features. In Appendix~\ref{section_Appendix_vLOS_vDopp}, we find that the LCT flows are weaker in regions within \SI{2}{\degree} of sunspots, compared to direct SDO/HMI Doppler velocity measurements. This difference can however be attributed to physical causes as well, that is, the difference in the height of the two measurements.

To ensure that our measurements are not susceptible to this unreliability, we create additional data cubes of normalized continuum intensity for each AR, by binning to the same grid spacing as the flow maps (\SI{0.4}{\degree}). We identify pixels which have values lower than 0.95 (pores, sunspots) and higher than 1.05 (plage), where the mean value of the quiet Sun is 1. In the flow maps, we exclude these pixels if they are connected to at least one other, to avoid excluding random high-valued pixels. The numbers are the result of trial and error, and represent a compromise between excluding pixels in the quiet Sun which are above/below the thresholds, and not excluding pixels which are systematically lower/higher, for example near the edge of a sunspot.

\subsection{Ensemble averages}
\label{subsec_averaging}

\begin{figure}
\centering
\includegraphics[width=\hsize]{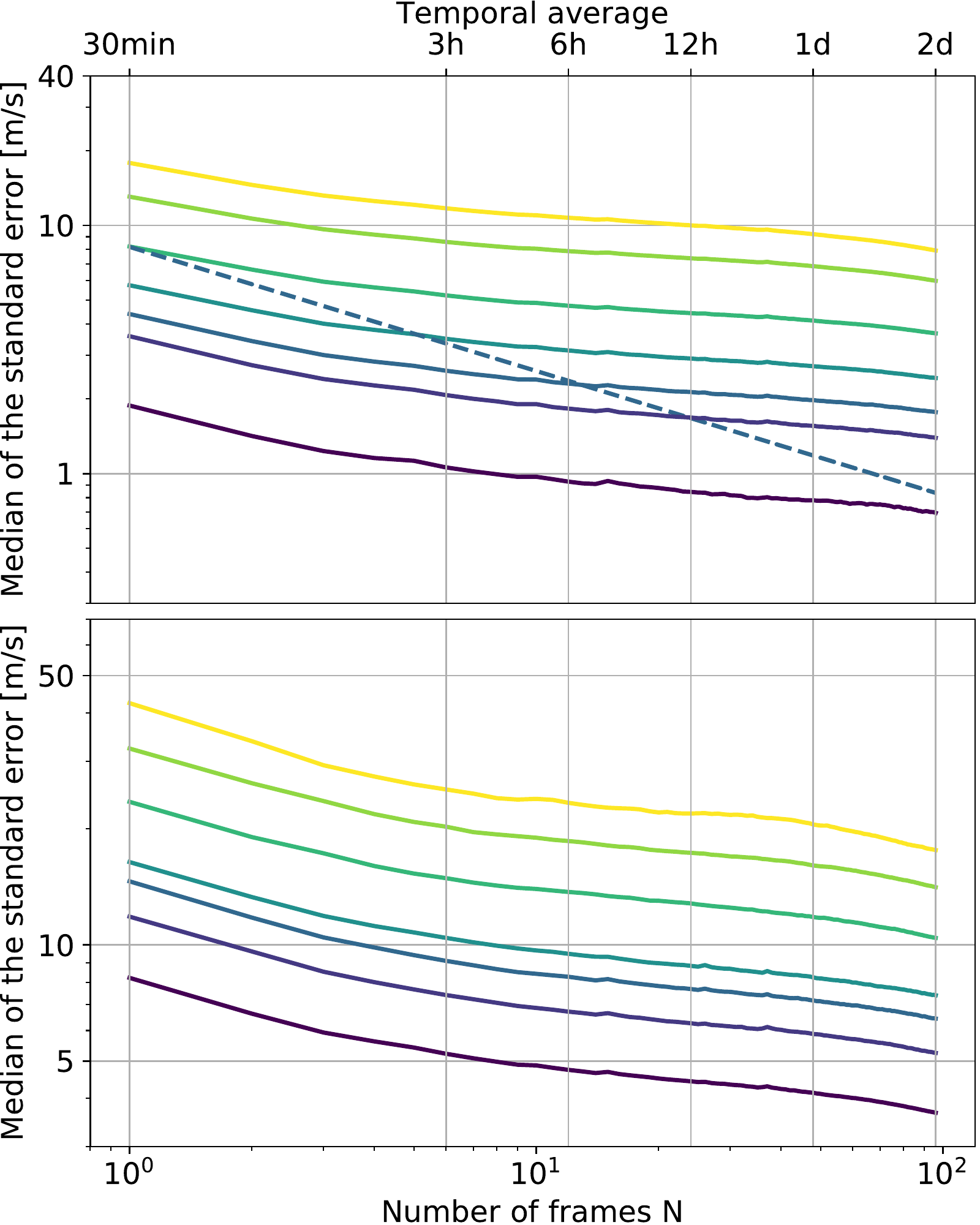}
  \caption{Top: Median of the standard error against averaging time. The lines show the cases of no smoothing (yellow) and of Gaussian smoothing of \SI{0.4}, \SI{0.8}, \SI{1.2}, \SI{1.6}, \SI{2.0} and \SI{4.0}{\degree} (green to purple). All lines are from an ensemble average over all 182~ARs. The dotted line shows the $1/\sqrt{N}$ least-square fit to the case of Gaussian smoothing of $\sigma = \SI{1.6}{\degree}$. Bottom: Same as top, for averages over 5, 10, 20, 45, 60, 91 and 182~control regions (yellow to purple). All lines are with a Gaussian smoothing of $\sigma= \SI{0.8}{\degree}$. Both panels show the case of the latitudinal flow component. The scale of the y-axis is not the same in the two panels.
	  }
      \label{Stderrplot}
\end{figure}

To carry out ensemble averages, we first investigate the dependence of the noise in the flow maps on spatial smoothing, temporal averaging, and the number of ARs in the ensemble average, to identify parameters that yield appropriate noise levels for our analysis. Previous studies reported active region inflows with velocities on the order of several tens of m~s${^{-1}}$. We therefore want the noise to be well below \SI{10}{\meter \per \second}, while still being able to investigate flows on small spatial scales and without losing much temporal resolution.

For this, we use the sample of control regions of the HEAR survey, where no systematic flows are present. The data cubes of these are prepared in the same way as those of the EARs, with the center of the \SI{60x60}{\degree} maps on the coordinate given in the HEAR survey throughout the observations. For this test, we crop the frames to the inner \SI{40x40}{\degree}, which yields a more representative value, since the errors increase towards the limb and fewer control regions contribute to the average. Like the active regions, the quiet Sun regions have different disk positions at a given time relative to the assigned 'time of emergence'. For a representative estimate, we start at one day before the time of emergence, where the frames from the control regions will on average be close to disk center, and compute the average of $N$ subsequent frames, in the range of $N=1$ (no time average, 30~minute resolution) to $N=96$ (2~days average). These are additionally smoothed by Gaussians of $\sigma = 0.4,0.8,1.2,1.6,2.0$ and \SI{4.0}{\degree}. We calculate the median of the standard error over all individual control regions.
Secondly, we carry out the same calculation, at a fixed Gaussian smoothing of \SI{0.8}{\degree}, while varying the number of control regions in the average. Fig.~\ref{Stderrplot} shows the results for the latitudinal flow component. The results for the longitudinal component are very similar.

The top panel of Fig.~\ref{Stderrplot} shows that an ensemble average over all 182~ARs requires a spatial smoothing of \SI{0.8}{\degree} to obtain a noise level on the order of \SI{5}{\meter \per \second}. A time average of six~hours or twelve~hours is sensible because of the time scale on which the flows change. The top panel also shows a least-square fit of $1 / \sqrt{N}$ to the case of Gaussian smoothing of $\sigma = \SI{1.6}{\degree}$. The noise decreases slower than the fit, because subsequent maps are correlated due to the lifetime of supergranules. This holds for all cases. The bottom panel of Fig.~\ref{Stderrplot} shows that for an ensemble average of fewer than 45~regions, temporal averaging of more than 2~days is necessary at this level of spatial smoothing. For an ensemble average of 45~regions, a time average of 12~hours is acceptable.

We use these values for spatial smoothing and temporal averaging for the ensemble averages in our analysis. The temporal averaging is aligned relative to the time of emergence, such that the frame of emergence is the first frame in the first time step after emergence. For the ensemble averaging, Hale's law and Joy's law are taken care of by reversing the latitudinal direction of all data cubes (magnetic field, continuum intensity and both $v_{\rm lon}$ and $v_{\rm lat}$ flow components), and the sign of the magnetogram and the $v_{\rm lat}$ flow component, for regions in the southern hemisphere. 

In addition to the flows around active regions, the flow maps exhibit signatures from the moat flow, due to the presence of sunspots in the sample \citep{Sheeley_1972}. To better understand their impact on the flow maps, we perform ensemble averages centered on the leading polarity (where we expect the strongest moat flow), in subsamples of the active regions, using a classification scheme to identify whether there is a sunspot with a clear penumbra present in an AR at that time or not. Appendix \ref{section_Appendix_SSQ} describes this test. We find that a moat flow is present in our flow data for regions which have a clear sunspot, with typical velocities of about \SI{150}{\meter \per \second}. The moat flow is not symmetric in this early phase of the active region, as the sunspots still grow, in agreement with \citet{Vargas-Dominguez_2008}.

\section{Results}
\label{Sec_Results}

\begin{figure}
\centering
\includegraphics[width=\hsize]{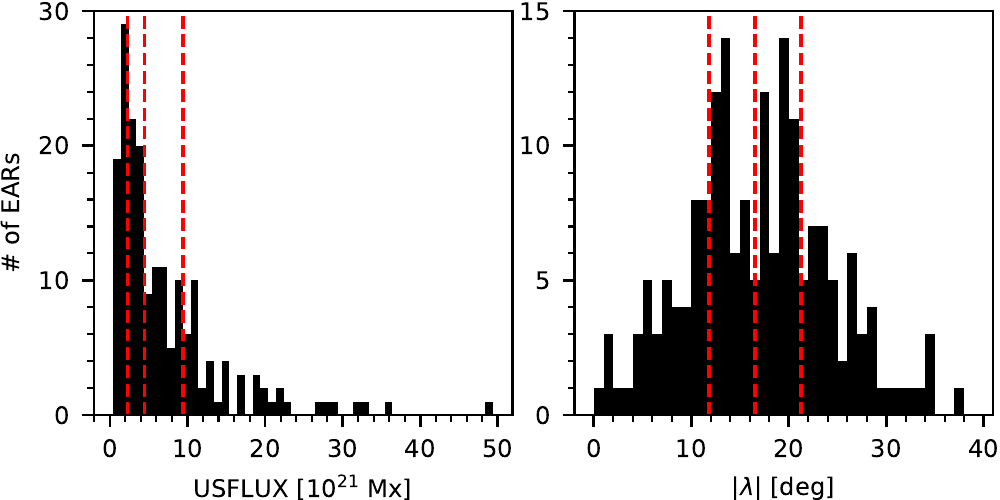}
  \caption{Left: Distribution of the unsigned magnetic flux (USFLUX) of the EARs. Right: Distribution of the unsigned latitude $|\lambda|$ of the EARs. In both plots, the red-dotted lines separate the distribution into four subsamples, sorted by the respective parameter, each containing 45 or 46 active regions.
	  }
      \label{subsample_Latitudes_histogram}
\end{figure}

\subsection{Flows as a function of time and magnetic flux}
\label{subsection_subsample_USFLUX}

\begin{figure*}
\centering
    {\includegraphics[width=1\textwidth]{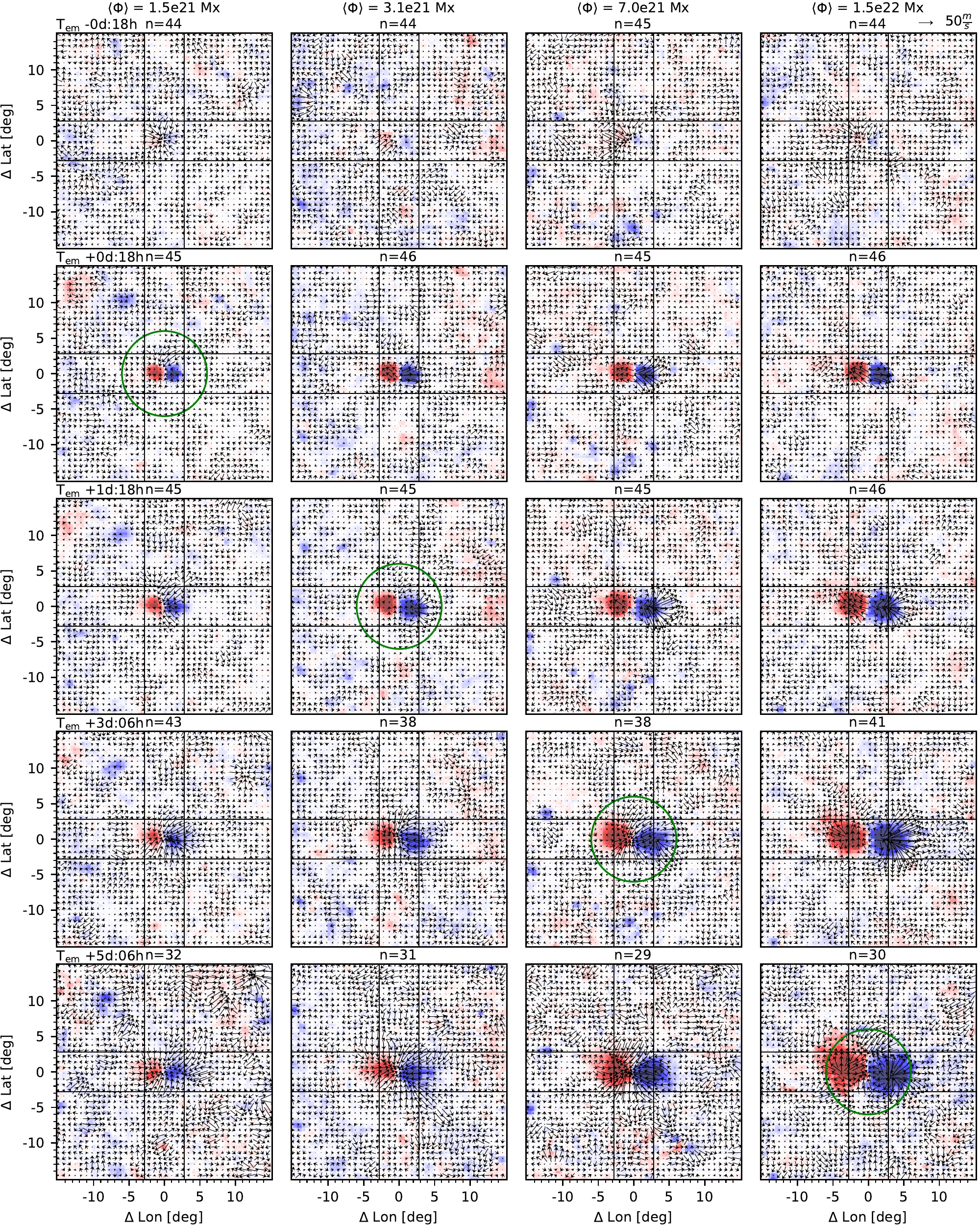}}
    \caption{
    Evolution (from top to bottom) of the averaged active region magnetic field and flows, for the four subsample averages of total unsigned flux (with flux increasing from left to right, cf. left panel of Fig.~\ref{subsample_Latitudes_histogram}). Each time step is an average over 12~hours, centered on the labeled time. The axis labels are relative to the center of the AR, as defined in Sect.~\ref{Sec_Positions}. For each panel, the total number of active regions that contribute to the average is given on top. Red (blue) indicates positive (negative) radial magnetic field, saturated at $\pm$\SI{150}{\Gauss}. The green circles highlight the time after emergence at which the inflows are first visible in the time steps shown here. The arrows indicate the flows. A reference arrow representing a velocity of \SI{50}{\meter \per \second} is given in the upper right corner. The black lines mark the averaging range which is used for the model fit (cf. text and Fig.~\ref{plot_modelfit_example}). A moat flow is apparent in for example the rightmost column at 3~days and 6~hours.
    }
  \label{subsample_USFLUX_plot}
\end{figure*}

We investigate the evolution of the flows around active regions of different maximum values of magnetic flux. For each of the 182~ARs, we retrieve the maximum total unsigned flux from the hmi.sharp\_720s series keyword USFLUX, in the period covered by the HEAR survey. USFLUX records $\Sigma |B_z| \mathrm{d}A$, with $B_z$ the radial magnetic field component and $A$ the area of the SHARP region \citep{Bobra_2014}. From this, we divide the sample into four subsamples, sorted by the maximum unsigned magnetix flux, with two subsamples containing 45 and two subsamples containing 46~EARs.

The left panel of Fig.~\ref{subsample_Latitudes_histogram} shows the distribution of maximum USFLUX values of all ARs, together with red-dotted lines indicating the boundaries between the four subsamples. The subsamples have mean total unsigned magnetic flux values of \SI{1.5e21}{Mx}, \SI{3.1e21}{Mx}, \SI{7.0e21}{Mx} and \SI{1.5e22}{Mx}, respectively.

Fig.~\ref{subsample_USFLUX_plot} shows several twelve-hour time steps for all four subsamples. The first row shows the time step of the flows at $-18$~hours (between 24~and 12~hours before emergence). At this time, all subsamples exhibit a converging flow towards the center of the AR. The velocities are around \SIrange{20}{30}{\meter \per \second}. The noise level is about \SI{10}{\meter \per \second}, as discussed in Sect.~\ref{subsec_averaging}. This converging flow is consistent with \citet{Birch_2019}. For the two higher-flux subsamples, the converging flow is spatially more extended than for the lower-flux ones. Shortly after emergence, the lowest-flux subsample still shows small-scale inflows, whereas they have ceased for the other subsamples. The three subsamples with higher flux show a prograde flow with velocities of about \SI{40}{\meter \per \second} at the position of the leading polarity, similar to the findings by \citet{Birch_2019}. This feature is not present in the lowest-flux subsample. About two~days after emergence (third row in Fig.~\ref{subsample_USFLUX_plot}), inflows towards the second-lowest flux subsample set in. For the third subsample, they are clearly visible after about three~days, while the highest-flux subsample still shows mostly a strong moat flow at this time. For the highest-flux subsample, the inflows only set in at around four~days after emergence. From this, we conclude that for large active regions, the inflows take longer to form.

After six~days, the EARs of the lowest-flux subsample have already mostly decayed. For these small AR, inflows are seen throughout their lifetime, and decay along with the magnetic field. Over its lifetime, the lowest-flux subsample never exhibits diverging flows, neither moat flows (due to the lack of sunspots with clear penumbra), nor a prograde flow at the leading polarity of the kind observed by \citet{Birch_2019}.

\subsection{Flows as a function of time and latitude}
\label{subsection_subsample_latitudes}

\begin{figure*}
\centering
    {\includegraphics[width=1\textwidth]{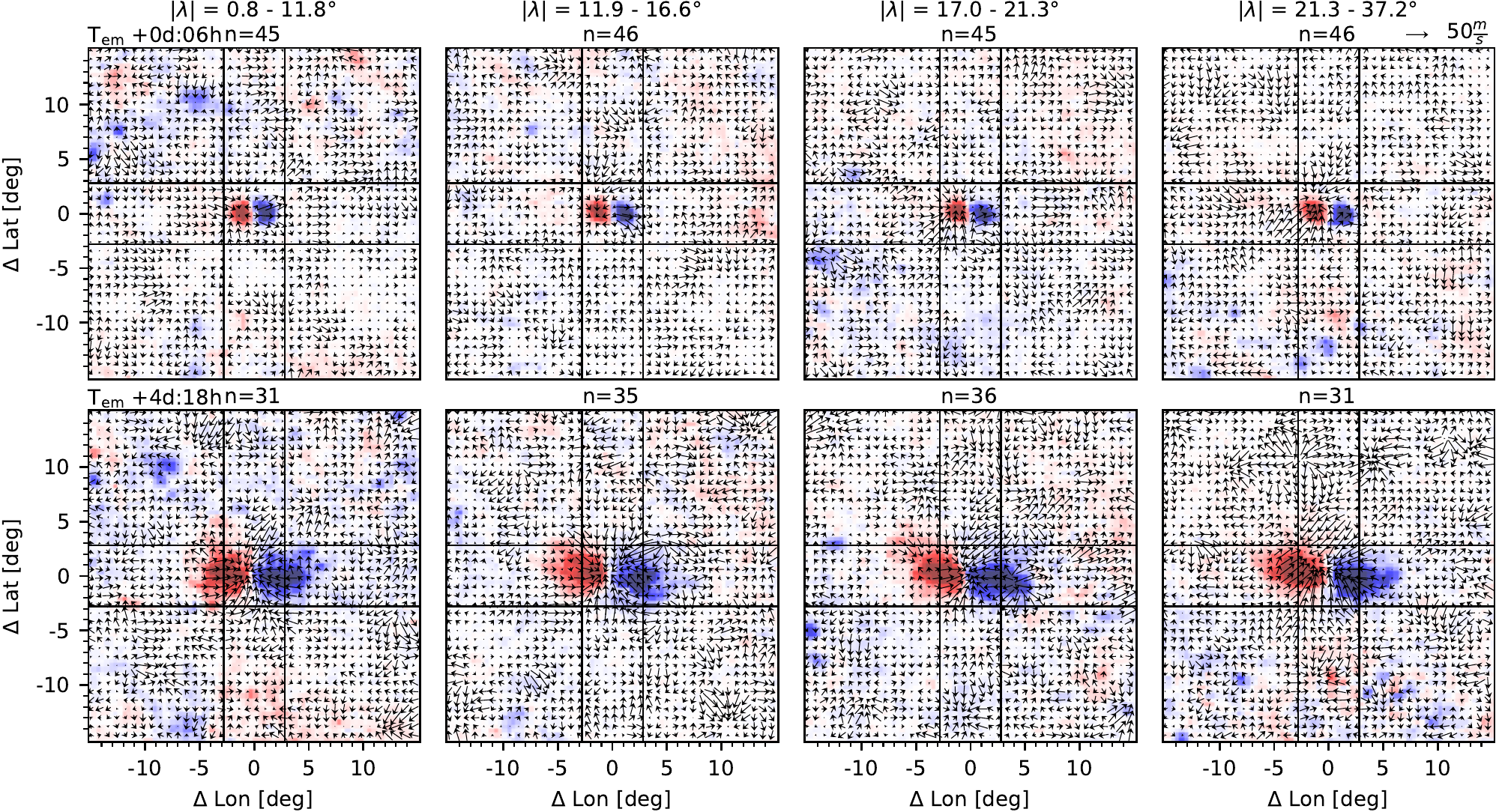}}
    \caption{Evolution (from top to bottom) of the averaged active region magnetic field and flows, for the four subsample averages of unsigned latitude $|\lambda|$ (with $|\lambda|$ increasing from left to right, cf. right panel of Fig.~\ref{subsample_Latitudes_histogram}). The labels, color scales and arrow scales are the same as in Fig.~\ref{subsample_USFLUX_plot}. The diverging flow in the bottom row of the third subsample is due to a strong moat flow in AR~11158, which is the largest AR in this subsample at this time.}
  \label{subsample_latitude_plot}
\end{figure*}

We examine whether the flows associated with the EARs depend on the latitude at which the EARs emerges. For that, we divide the sample of 182~ARs into four subsamples of 45 or 46~EARs each, sorted by their unsigned latitude $| \lambda |$ (as provided by the HEAR survey). Each subsample in $| \lambda |$ therefore contains regions with a variety of maximum total unsigned flux values.

The right panel of Fig.~\ref{subsample_Latitudes_histogram} shows the distribution of unsigned latitudes of all EARs, together with lines indicating the boundaries between the four subsamples.

Fig.~\ref{subsample_latitude_plot} shows two twelve-hour time steps for all four subsamples, at six~hours after emergence and 4 days and 18 hours after emergence. The converging flow towards the center of the active region as well as the prograde flow at the position of the leading polarity around the time of emergence are both present in all four subsamples.
Joy's law is clearly visible after emergence, consistent with \citet{Schunker_2020} who show that active regions emerge on average east-west aligned, and that Joy's law becomes evident two days after emergence. The tilt angle increases towards the higher-latitude subsamples and in time. There appears to be no striking systematic change in the inflows as a function of latitude.

\subsection{Flows averaged over the full sample as a function of time}
While the sections above outline the evolution of the flows in subsamples of active regions with respect to flux and latitude, we want to put this into the context of the average EAR over the whole sample. Appendix~\ref{section_Appendix_fullensemble} provides additional plots of the average flows around the time of emergence as well as averages in longitude covering the leading and the trailing polarity separately.
Overall, the full ensemble average again exhibits converging flows starting one day prior to emergence. The inflows have a maximum extent of about \SI{7}{\degree} from the center of the AR, in the time between emergence and seven~days after emergence, and amplitudes of about \SI{50}{\meter \per \second}, at a noise level of about \SI{5}{\meter \per \second}.

\subsection{Quantitative model of the inflows as a function of time and magnetic flux}

\begin{figure}
\centering
\includegraphics[width=\hsize]{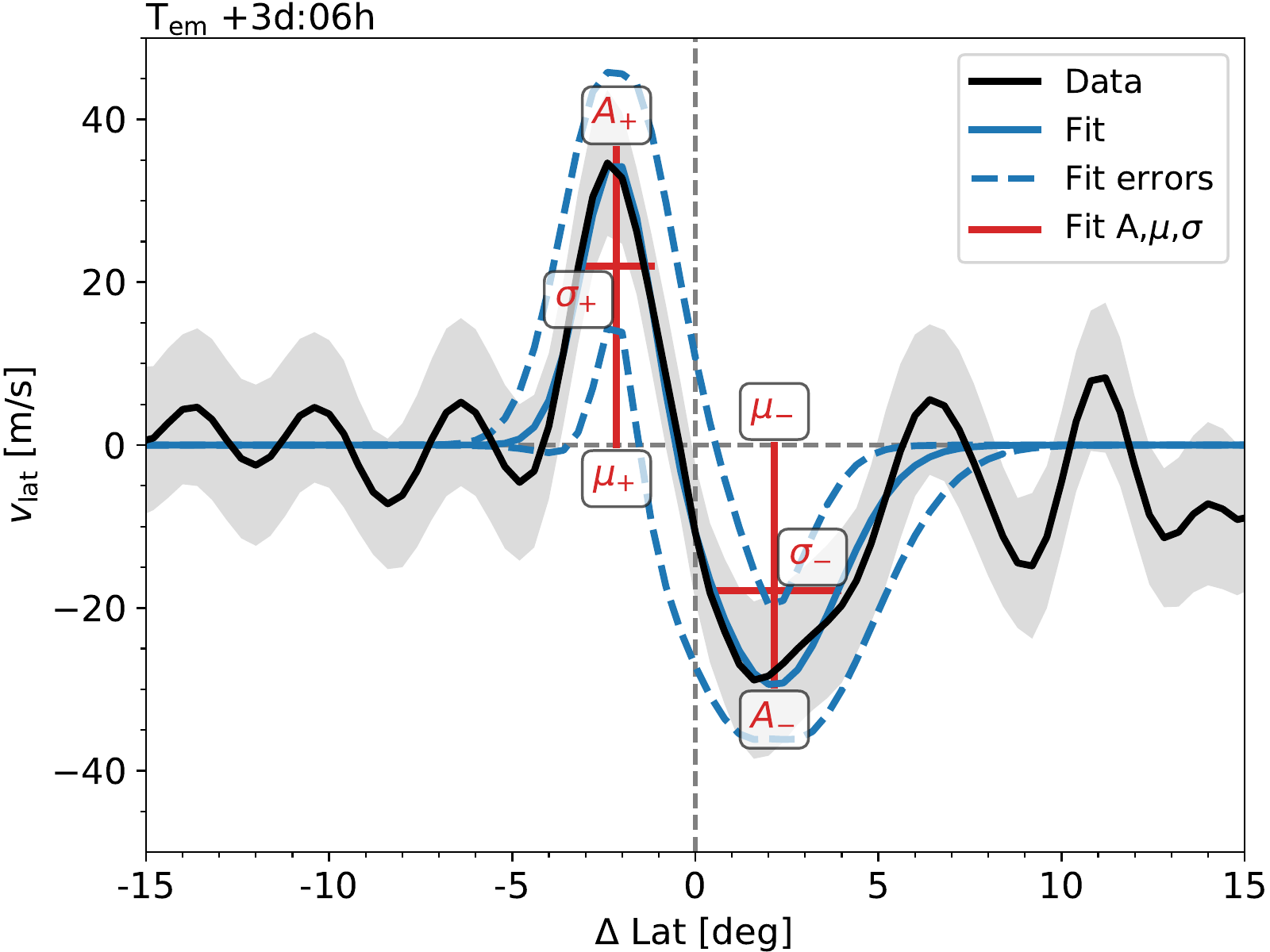}
  \caption{
    Data and model fit of a time step of the second-highest flux subsample ($\langle \Phi \rangle=\SI{7.0e21}{\Maxwell}$). The black line shows the latitudinal flow, averaged over longitudes in the range indicated in Fig.~\ref{subsample_USFLUX_plot}. The solid blue line shows the model fit of two Gaussians with opposite polarity (see text for details). The dashed blue lines show the $1 \sigma$ error estimates of the fit.
    The red markings indicate the fit parameters of the Amplitude $A$ (in \SI{}{\meter \per \second}), the peak position $\mu$ (relative to the center of the active region along latitude, $\Delta$ Lat), and the standard deviation $\sigma$ (in degree) of both Gaussians. The gray shaded region indicates the rms of the flow data. The fit parameters of the two Gaussians slightly deviate from the fitted curve due to superposition; this is however small compared to the error.
    }
      \label{plot_modelfit_example}
\end{figure}

\begin{figure*}
\centering
\includegraphics[width=\hsize]{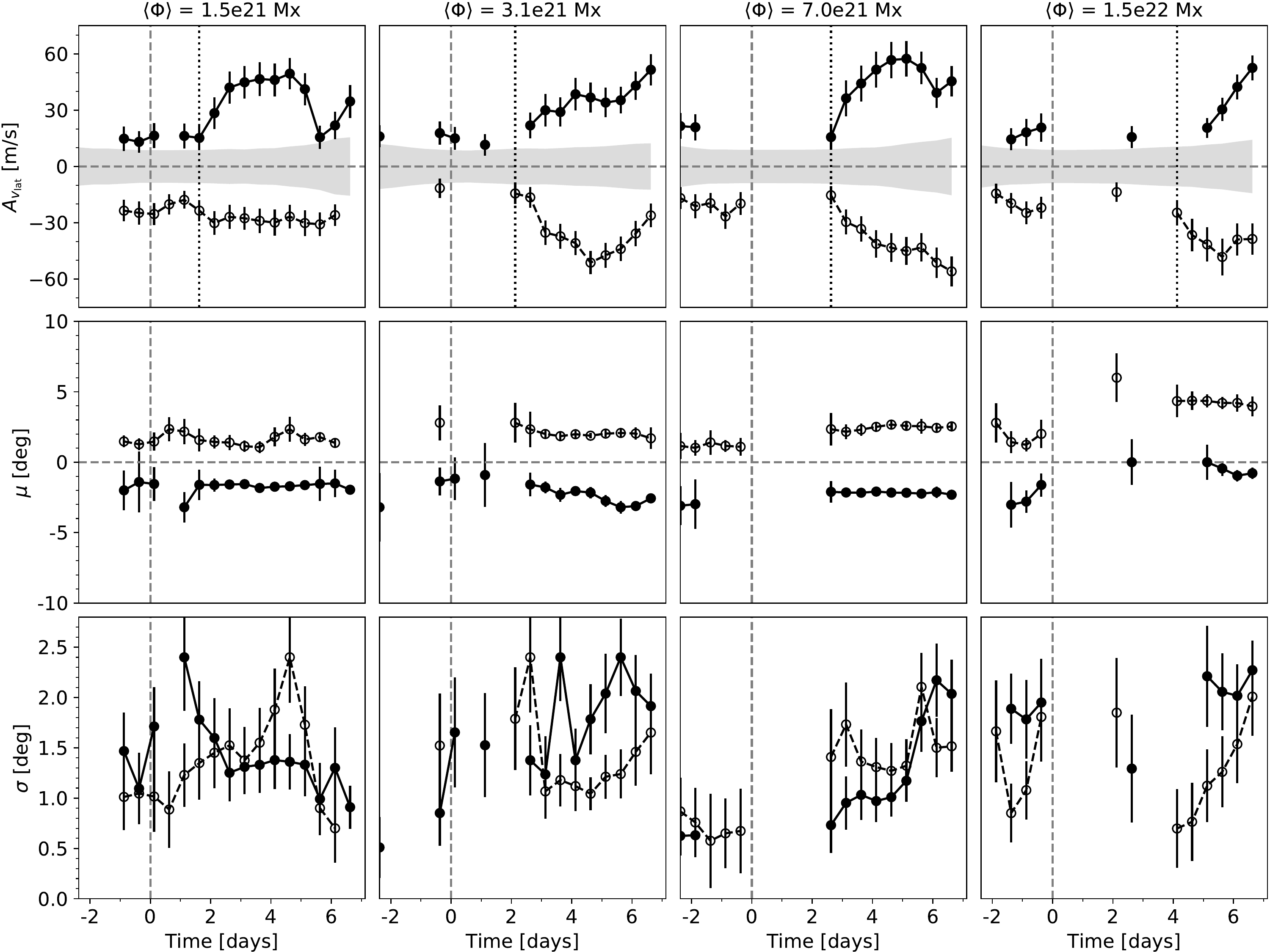}
  \caption{Parameters of the best-fit Gaussians to the longitudinally averaged latitudinal flow (see Fig.~\ref{plot_modelfit_example}). The top row shows the amplitudes $A_+$ and $A_-$, the middle row shows the peak positions $\mu_+$ and $\mu_-$, and the bottom row shows the standard deviations $\sigma_+$ and $\sigma_-$ of the model fits (c.f. Fig.~\ref{plot_modelfit_example}). The four columns show the averages over different ranges of flux, with flux increasing from left to right. The error bars are Monte Carlo estimates using quiet Sun control regions as background noise. No model is fitted when there are no inflow signatures above the noise level (see text). The filled dots (circles) and solid (dashed) lines indicate poleward (equatorward) velocity. The dotted black lines indicate the onset time of the latitudinal flow for each subsample. The gray shaded region indicates the rms in the quiet Sun control regions.
	  }
  \label{plot_1Dfit_Fitparameters_vlat_along_lon}
\end{figure*}

\begin{figure}
\centering
\includegraphics[width=\hsize]{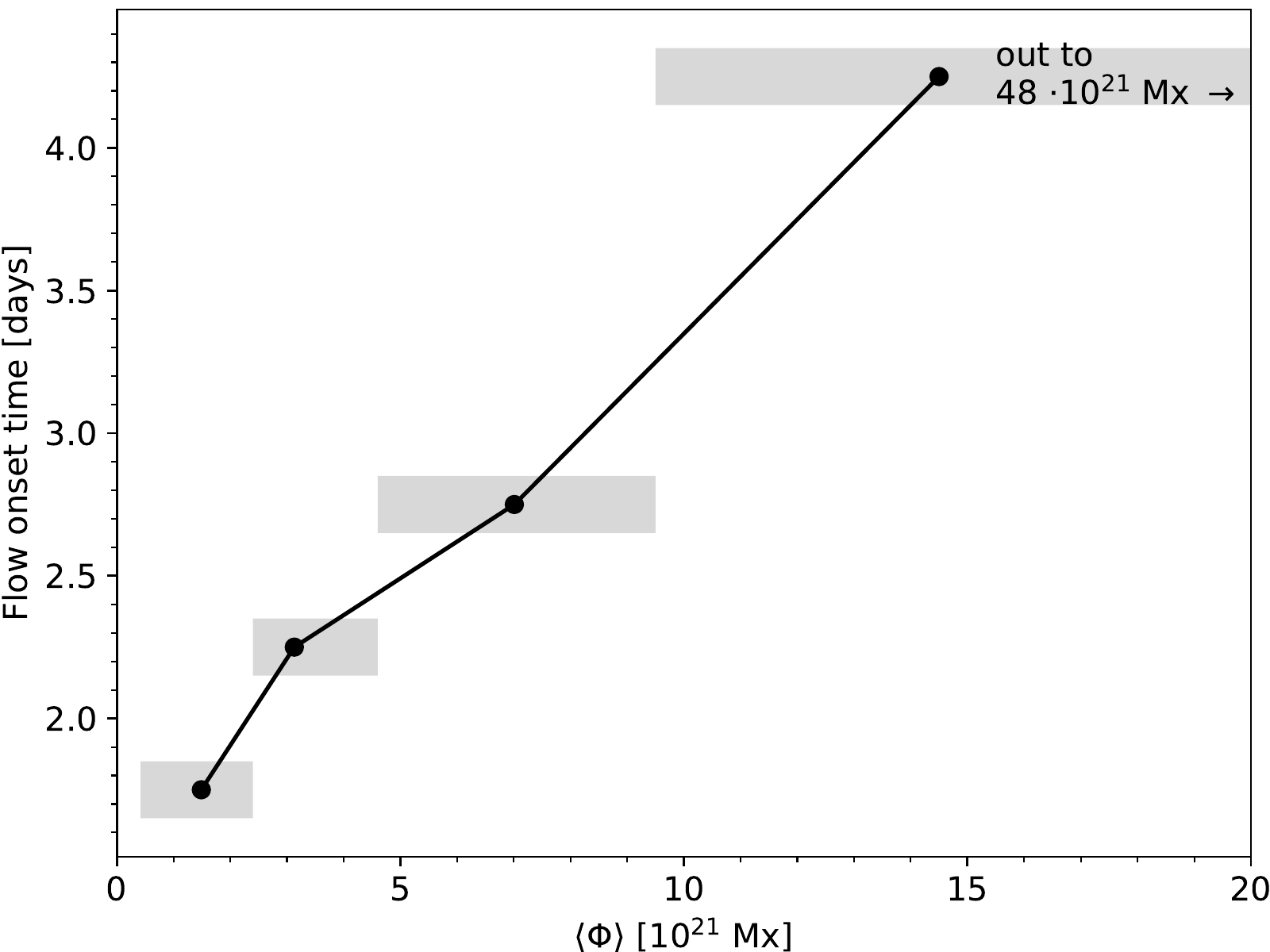}
  \caption{Flow onset times in relation to mean flux. The gray boxes indicate the flux ranges of the subsamples. The plot is truncated at \SI{2e22}{\Maxwell}; the range for the highest-flux subsample continues to \SI{4.8e22}{\Maxwell}.
	  }
  \label{plot_flowonset_vs_flux}
\end{figure}

We want to make a quantitative description of the inflows around active regions and to evaluate the differences found in the flux-binned subsamples (Sect.~\ref{subsection_subsample_USFLUX}). For this, we fit a model to the flows. The main contribution of the inflows is along the latitudinal axis. We therefore average the latitudinal flows over the central \SI{5.6}{\degree} in longitude for each time step and each subsample. This range is chosen to cover a large part of the AR, while excluding most of the moat flow signature that is present in the high-flux subsamples. We do not attempt to fit a separate model to the moat flow, as the sunspots and with them the moat flow in the individual active regions form at different times relative to the time of emergence, and in some cases break up before the end of the observations (cf. the lower panels for the second-highest flux subsample in Fig.~\ref{subsample_USFLUX_plot}).

The model we apply consists of the sum of up to two 1D Gaussians with positive/negative amplitude $A$, position $\mu$ and width $\sigma$. As an estimate of the background noise, we compute the standard deviation over the same \SI{5.6}{\degree} in longitude, for subsamples of the control regions corresponding to the AR subsamples. If the amplitude of a fitted Gaussian is smaller than 1.3~times the noise, it is discarded. The converging flows before emergence tend to be less spatially extended than the inflows after emergence. We therefore confine the region of the fits before emergence to about \SI{2}{\degree} in latitude, in order to avoid fitting of large supergranules.

To estimate realistic errors on the fit parameters, we add the model fit onto the control region frames and rerun the model fitting on these. The control region flow maps have the same properties (spatial scale, amplitude) as the active region flow maps, and provide a background noise that is uncorrelated with the background in the active region maps. Averaging over all realizations of this procedure and taking the standard deviation of each parameter from these realizations yields an error estimate for the model parameters. Of the 30~time steps of the control region subsamples, the first and last five are not used because of their proximity to the limb, which would artificially increase the noise. Thus, for each model fit, there are $4 \times 20 = 80$~realizations that contribute to the error estimate. Fig.~\ref{plot_modelfit_example} shows an example of the model fit. The calculated errors (blue dashed lines) are of the same order as the noise in the data (gray shaded region).

Fig.~\ref{plot_1Dfit_Fitparameters_vlat_along_lon} shows the fit parameters $A$, $\mu$ and $\sigma$ of $v_{\rm lat}$ averaged over longitudes, for the four flux-binned subsamples, in the time range between one~day before and six~days after emergence (the last time step is left out because of low signal to noise). For the three higher-flux subsamples, the converging flows of the pre-emergence phase vanish shortly after emergence, such that there is no successful fit. After some days, the inflows set in. We define the onset time of the inflows as the time when the model fits start to be continuously successful. The amplitudes increase from this time on, starting from less than \SI{20}{\meter \per \second} and reaching maximum velocities of \SIrange{50}{60}{\meter\per\second} in all cases. In the case of the lowest-flux subsample, where there is no gap in the model fits, we define the onset time as the time when the amplitude of the fits starts to increase, similar to the other subsamples.

To verify that the moat flow does not affect these measurements, we performed the same analysis again, excluding those pixels from the fit which lie within the area of the moat flow for more than half of the ARs in the respective subsample. We approximate the moat flow area as regions of \SI{2}{\degree} surrounding positions with reduced continuum intensity (see Sect.~\ref{subsec_blankout} and Appendix~\ref{section_Appendix_vLOS_vDopp}). This procedure leaves the model fits and onset times unaffected, except for the positive-valued Gaussian in the highest-flux subsample, where only that part of the inflow that lies outside of the central region is fitted, resulting in reduced amplitude and width of the fit.

The onset times illustrate what was discussed qualitatively in Sect.~\ref{subsection_subsample_USFLUX}: The inflows towards the center of the active region set in at increasingly later times for increasing amount of flux of the AR, in the range between one and four days after emergence. Fig.~\ref{plot_flowonset_vs_flux} shows the onset time against the mean flux of each subsample. We point out that the subsamples differ substantially in flux range (cf. left panel of Fig.~\ref{subsample_USFLUX_plot}).

We measure the extent of the inflows from the models as the peak position $\mu$ added to the Half Width at Half Maximum (HWHM) of the fit. With this, we find inflow extents of about \SIrange{5}{7}{\degree} with an error of about \SI{1}{\degree}.
After five~days, the inflows around the lowest-flux subsample decrease in width and amplitude. For the other subsamples, the fits illustrate that at the end of the observed time period, the flows have not reached a steady state, especially for the highest-flux subsample, where the inflows after emergence only formed two~days before the end of the observational period.

\section{Discussion}
\label{Sec_Discussion}

We investigated the evolution of surface flows associated with active regions before, during and up to six~days after their emergence to the surface, using Local Correlation Tracking of granulation on the HEAR survey of 182~emerging active regions. We tested the processing of the flow measurements with synthetic input data and by cross-correlation with independent measurements from direct Doppler images.

About half of the active regions in the sample develop a sunspot with a clear penumbra in the observed period, which show moat flow structures with velocities on the order of \SI{150}{\meter \per \second} in the prograde and equator- and poleward direction.

To study the evolution of the inflows around active regions, we carried out ensemble averages in subsamples ordered by total unsigned magnetic flux and unsigned latitude of the active regions as well as an average over all regions. We then fitted a model to the latitudinal flow component in the subsamples ordered by flux.

We find that AR emergence is preceded by converging flows of around \SIrange{20}{30}{\meter \per \second}, beginning one day before emergence. This agrees with results reported by \citet{Birch_2019}, and holds for all ARs in our sample, that is, it is independent of the maximum flux or the latitude of the AR. However, the extent of these early inflows is larger for stronger ARs.

We find that these pre-emergence converging flows cease shortly after emergence. Larger-scale latitudinal inflows form in the days following emergence. The time at which they form depends on the flux of the AR, between about one and four~days, and increases with flux. There are several possibilities for the cause of this effect: It could be related to the separation speed of the polarities. \citet{Schunker_2019} identified different phases of emergence with respect to the separation speed, finding that on average over all AR in the HEAR sample, the separation stops increasing at 2.5~-~3~days after emergence. However, the distinction they make between regions with flux higher or lower than the median value indicates that this time also depends on the flux. It is therefore possible that the onset of the inflow coincides with the end of the separation increase. Another possibility is that it could depend on the relative amount of emerged flux, such that the inflows set in at a certain time relative to the peak flux. It could also depend on the ratio of flux within sunspots vs. flux within plage, since the proposed mechanism for driving the inflows depends on the plage \citep{Spruit_2003}. Further investigation will be necessary to clarify this.

Our ensemble averages show a maximum extent of the inflows of around \SI{7}{\degree}, which is somewhat smaller than the \SI{10}{\degree} reported recently by \citet{Braun_2019}. Typical velocities during the later stages are around \SI{50}{\meter \per \second}, which is at the high end of previously reported velocities \citep{Gizon_2001, Haber_Hindman_2004, Loeptien_2017, Braun_2019}. This is in part attributable to differences in spatial resolution; with lower resolution (by e.g. spatial smoothing with a broader Gaussian), the inflows would have lower amplitudes and larger extents. Also, our sample consists of emerging active regions, whereas previous studies analyzed well-established, long-lived active regions.

Our measurements of the flows show that they are still evolving at the end of the observed time frame. Observations covering the transition to a steady state are needed to shed more light on the physical context of the flows. Since the observation of the emergence is necessary for an accurate age determination of an active region, the limitation to seven~days before and after emergence is a constraint which is at present difficult to circumvent. One possibility is to add observations from different heliographic longitudes, from for example the recently launched Solar Orbiter, which increases the time that an active region can be tracked continuously. Another option is to track the ARs from the HEAR sample to the next rotation. This is however limited to the very long lived regions, and leaves an observational gap of more than two weeks.

\begin{acknowledgements}
N.G. is a member of the International Max Planck Research School (IMPRS) for Solar System Science at the University of Göttingen.
N.G. conducted the data analysis, contributed to the interpretation of the results, and wrote the manuscript.
We thank Jesper Schou, Paul-Louis Poulier and Bastian Proxauf for useful discussions.
The HMI data used here are courtesy of NASA/SDO and the HMI Science Team.
We acknowledge partial support from the European Research Council Synergy Grant WHOLE SUN \#810218.
The data were processed at the German Data Center for SDO, funded by the German Aerospace Center under grant DLR50OL1701.
This research made use of Astropy,\footnote{http://www.astropy.org} a community-developed core Python package for Astronomy \citep{astropy:2013, astropy:2018}. This work used the NumPy \citep{oliphant2006guide}, SciPy \citep{2020SciPy-NMeth}, pandas \citep{mckinney-proc-scipy-2010} and Matplotlib \citep{Hunter:2007} Python packages.

\end{acknowledgements}


\bibliographystyle{aa.bst}
\bibliography{Jabref_PaperI.bib}

\begin{thebibliography}{59}
\expandafter\ifx\csname natexlab\endcsname\relax\def\natexlab#1{#1}\fi

\bibitem[{{Astropy Collaboration} {et~al.}(2018){Astropy Collaboration},
  {Price-Whelan}, {Sip{\H{o}}cz}, {G{\"u}nther}, {Lim}, {Crawford}, {Conseil},
  {Shupe}, {Craig}, {Dencheva}, {Ginsburg}, {VanderPlas}, {Bradley},
  {P{\'e}rez-Su{\'a}rez}, {de Val-Borro}, {Paper Contributors}, {Aldcroft},
  {Cruz}, {Robitaille}, {Tollerud}, {Coordination Committee}, {Ardelean},
  {Babej}, {Bach}, {Bachetti}, {Bakanov}, {Bamford}, {Barentsen}, {Barmby},
  {Baumbach}, {Berry}, {Biscani}, {Boquien}, {Bostroem}, {Bouma}, {Brammer},
  {Bray}, {Breytenbach}, {Buddelmeijer}, {Burke}, {Calderone}, {Cano
  Rodr{\'\i}guez}, {Cara}, {Cardoso}, {Cheedella}, {Copin}, {Corrales},
  {Crichton}, {D{\textquoteright}Avella}, {Deil}, {Depagne}, {Dietrich},
  {Donath}, {Droettboom}, {Earl}, {Erben}, {Fabbro}, {Ferreira}, {Finethy},
  {Fox}, {Garrison}, {Gibbons}, {Goldstein}, {Gommers}, {Greco}, {Greenfield},
  {Groener}, {Grollier}, {Hagen}, {Hirst}, {Homeier}, {Horton}, {Hosseinzadeh},
  {Hu}, {Hunkeler}, {Ivezi{\'c}}, {Jain}, {Jenness}, {Kanarek}, {Kendrew},
  {Kern}, {Kerzendorf}, {Khvalko}, {King}, {Kirkby}, {Kulkarni}, {Kumar},
  {Lee}, {Lenz}, {Littlefair}, {Ma}, {Macleod}, {Mastropietro}, {McCully},
  {Montagnac}, {Morris}, {Mueller}, {Mumford}, {Muna}, {Murphy}, {Nelson},
  {Nguyen}, {Ninan}, {N{\"o}the}, {Ogaz}, {Oh}, {Parejko}, {Parley}, {Pascual},
  {Patil}, {Patil}, {Plunkett}, {Prochaska}, {Rastogi}, {Reddy Janga},
  {Sabater}, {Sakurikar}, {Seifert}, {Sherbert}, {Sherwood-Taylor}, {Shih},
  {Sick}, {Silbiger}, {Singanamalla}, {Singer}, {Sladen}, {Sooley},
  {Sornarajah}, {Streicher}, {Teuben}, {Thomas}, {Tremblay}, {Turner},
  {Terr{\'o}n}, {van Kerkwijk}, {de la Vega}, {Watkins}, {Weaver}, {Whitmore},
  {Woillez}, {Zabalza}, \& {Contributors}}]{astropy:2018}
{Astropy Collaboration}, {Price-Whelan}, A.~M., {Sip{\H{o}}cz}, B.~M., {et~al.}
  2018, \aj, 156, 123

\bibitem[{{Astropy Collaboration} {et~al.}(2013){Astropy Collaboration},
  {Robitaille}, {Tollerud}, {Greenfield}, {Droettboom}, {Bray}, {Aldcroft},
  {Davis}, {Ginsburg}, {Price-Whelan}, {Kerzendorf}, {Conley}, {Crighton},
  {Barbary}, {Muna}, {Ferguson}, {Grollier}, {Parikh}, {Nair}, {Unther},
  {Deil}, {Woillez}, {Conseil}, {Kramer}, {Turner}, {Singer}, {Fox}, {Weaver},
  {Zabalza}, {Edwards}, {Azalee Bostroem}, {Burke}, {Casey}, {Crawford},
  {Dencheva}, {Ely}, {Jenness}, {Labrie}, {Lim}, {Pierfederici}, {Pontzen},
  {Ptak}, {Refsdal}, {Servillat}, \& {Streicher}}]{astropy:2013}
{Astropy Collaboration}, {Robitaille}, T.~P., {Tollerud}, E.~J., {et~al.} 2013,
  \aap, 558, A33

\bibitem[{{Birch} {et~al.}(2016){Birch}, {Schunker}, {Braun}, {Cameron},
  {Gizon}, {Lo ptien}, \& {Rempel}}]{Birch_2016}
{Birch}, A.~C., {Schunker}, H., {Braun}, D.~C., {et~al.} 2016, Science
  Advances, 2, e1600557

\bibitem[{{Birch} {et~al.}(2019){Birch}, {Schunker}, {Braun}, \&
  {Gizon}}]{Birch_2019}
{Birch}, A.~C., {Schunker}, H., {Braun}, D.~C., \& {Gizon}, L. 2019, \aap, 628,
  A37

\bibitem[{{Bobra} {et~al.}(2014){Bobra}, {Sun}, {Hoeksema}, {Turmon}, {Liu},
  {Hayashi}, {Barnes}, \& {Leka}}]{Bobra_2014}
{Bobra}, M.~G., {Sun}, X., {Hoeksema}, J.~T., {et~al.} 2014, \solphys, 289,
  3549

\bibitem[{{Bogart} {et~al.}(2011){Bogart}, {Baldner}, {Basu}, {Haber}, \&
  {Rabello-Soares}}]{Bogart_2011a}
{Bogart}, R.~S., {Baldner}, C., {Basu}, S., {Haber}, D.~A., \&
  {Rabello-Soares}, M.~C. 2011, in Journal of Physics Conference Series, Vol.
  271, GONG-SoHO 24: A New Era of Seismology of the Sun and Solar-Like Stars,
  012008

\bibitem[{{Brandenburg}(2005)}]{Brandenburg_2005}
{Brandenburg}, A. 2005, \apj, 625, 539

\bibitem[{{Braun}(2019)}]{Braun_2019}
{Braun}, D.~C. 2019, \apj, 873, 94

\bibitem[{{De Rosa} \& {Schrijver}(2006)}]{DeRosa_Schrijver_2006}
{De Rosa}, M.~L. \& {Schrijver}, C.~J. 2006, in ESA Special Publication, Vol.
  624, Proceedings of SOHO 18/GONG 2006/HELAS I, Beyond the spherical Sun, 12

\bibitem[{{Duvall}(1979)}]{Duvall_1979}
{Duvall}, T.~L., J. 1979, \solphys, 63, 3

\bibitem[{{Duvall} \& {Birch}(2010)}]{Duvall_2010}
{Duvall}, T.~L., J. \& {Birch}, A.~C. 2010, \apjl, 725, L47

\bibitem[{{Duvall} {et~al.}(1993){Duvall}, {Jefferies}, {Harvey}, \&
  {Pomerantz}}]{Duvall_1993}
{Duvall}, T.~L., J., {Jefferies}, S.~M., {Harvey}, J.~W., \& {Pomerantz}, M.~A.
  1993, \nat, 362, 430

\bibitem[{{Fan}(2009)}]{Fan_LRSP_2009}
{Fan}, Y. 2009, Living Reviews in Solar Physics, 6, 4

\bibitem[{{Fisher} \& {Welsch}(2008)}]{Fisher_Welsch_2008}
{Fisher}, G.~H. \& {Welsch}, B.~T. 2008, in Astronomical Society of the Pacific
  Conference Series, Vol. 383, Subsurface and Atmospheric Influences on Solar
  Activity, ed. R.~{Howe}, R.~W. {Komm}, K.~S. {Balasubramaniam}, \& G.~J.~D.
  {Petrie}, 373

\bibitem[{{Gary} \& {Hagyard}(1990)}]{Gary_1990}
{Gary}, G.~A. \& {Hagyard}, M.~J. 1990, \solphys, 126, 21

\bibitem[{{Giovanelli}(1980)}]{Giovanelli_1980}
{Giovanelli}, R.~G. 1980, \solphys, 67, 211

\bibitem[{{Gizon} \& {Birch}(2005)}]{Gizon_LRSP_2005}
{Gizon}, L. \& {Birch}, A.~C. 2005, Living Reviews in Solar Physics, 2, 6

\bibitem[{{Gizon} {et~al.}(2000){Gizon}, {Duvall}, \& {Larsen}}]{Gizon_2000}
{Gizon}, L., {Duvall}, T.~L., J., \& {Larsen}, R.~M. 2000, Journal of
  Astrophysics and Astronomy, 21, 339

\bibitem[{{Gizon} {et~al.}(2001){Gizon}, {Duvall}, \& {Larsen}}]{Gizon_2001}
{Gizon}, L., {Duvall}, Jr., T.~L., \& {Larsen}, R.~M. 2001, in IAU Symposium,
  Vol. 203, Recent Insights into the Physics of the Sun and Heliosphere:
  Highlights from SOHO and Other Space Missions, ed. P.~{Brekke}, B.~{Fleck},
  \& J.~B. {Gurman}, 189

\bibitem[{{Haber} {et~al.}(2004){Haber}, {Hindman}, {Toomre}, \&
  {Thompson}}]{Haber_Hindman_2004}
{Haber}, D.~A., {Hindman}, B.~W., {Toomre}, J., \& {Thompson}, M.~J. 2004,
  \solphys, 220, 371

\bibitem[{{Hathaway} {et~al.}(2002){Hathaway}, {Beck}, {Han}, \&
  {Raymond}}]{Hathaway_2002}
{Hathaway}, D.~H., {Beck}, J.~G., {Han}, S., \& {Raymond}, J. 2002, \solphys,
  205, 25

\bibitem[{{Hindman} {et~al.}(2009){Hindman}, {Haber}, \&
  {Toomre}}]{Hindman_2009}
{Hindman}, B.~W., {Haber}, D.~A., \& {Toomre}, J. 2009, \apj, 698, 1749

\bibitem[{{Hoeksema} {et~al.}(2014){Hoeksema}, {Liu}, {Hayashi}, {Sun},
  {Schou}, {Couvidat}, {Norton}, {Bobra}, {Centeno}, {Leka}, {Barnes}, \&
  {Turmon}}]{Hoeksema_2014}
{Hoeksema}, J.~T., {Liu}, Y., {Hayashi}, K., {et~al.} 2014, \solphys, 289, 3483

\bibitem[{Hunter(2007)}]{Hunter:2007}
Hunter, J.~D. 2007, Computing in Science \& Engineering, 9, 90

\bibitem[{{Jackiewicz} {et~al.}(2008){Jackiewicz}, {Gizon}, \&
  {Birch}}]{Jackiewicz_2008}
{Jackiewicz}, J., {Gizon}, L., \& {Birch}, A.~C. 2008, \solphys, 251, 381

\bibitem[{{Komm} {et~al.}(2011){Komm}, {Howe}, \& {Hill}}]{Komm_2011}
{Komm}, R., {Howe}, R., \& {Hill}, F. 2011, \solphys, 268, 407

\bibitem[{{Komm} {et~al.}(2012){Komm}, {Howe}, \& {Hill}}]{Komm_2012}
{Komm}, R., {Howe}, R., \& {Hill}, F. 2012, \solphys, 277, 205

\bibitem[{{Leighton}(1964)}]{Leighton_1964}
{Leighton}, R.~B. 1964, \apj, 140, 1547

\bibitem[{{Lindsey} \& {Braun}(2000)}]{Lindsey_2000}
{Lindsey}, C. \& {Braun}, D.~C. 2000, \solphys, 192, 261

\bibitem[{{Lisle} \& {Toomre}(2004)}]{Lisle_2004}
{Lisle}, J. \& {Toomre}, J. 2004, in ESA Special Publication, Vol. 559, SOHO 14
  Helio- and Asteroseismology: Towards a Golden Future, ed. D.~{Danesy}, 556

\bibitem[{{L{\"o}ptien} {et~al.}(2017){L{\"o}ptien}, {Birch}, {Duvall},
  {Gizon}, {Proxauf}, \& {Schou}}]{Loeptien_2017}
{L{\"o}ptien}, B., {Birch}, A.~C., {Duvall}, T.~L., {et~al.} 2017, \aap, 606,
  A28

\bibitem[{{L{\"o}ptien} {et~al.}(2016{\natexlab{a}}){L{\"o}ptien}, {Birch},
  {Duvall}, {Gizon}, \& {Schou}}]{Loeptien_2016_DataCompression}
{L{\"o}ptien}, B., {Birch}, A.~C., {Duvall}, T.~L., {Gizon}, L., \& {Schou}, J.
  2016{\natexlab{a}}, \aap, 587, A9

\bibitem[{{L{\"o}ptien} {et~al.}(2016{\natexlab{b}}){L{\"o}ptien}, {Birch},
  {Duvall}, {Gizon}, \& {Schou}}]{Loeptien_2016_shrinkingSun}
{L{\"o}ptien}, B., {Birch}, A.~C., {Duvall}, T.~L., {Gizon}, L., \& {Schou}, J.
  2016{\natexlab{b}}, \aap, 590, A130

\bibitem[{{L{\"o}ptien} {et~al.}(2018){L{\"o}ptien}, {Gizon}, {Birch}, {Schou},
  {Proxauf}, {Duvall}, {Bogart}, \& {Christensen}}]{Loeptien_2018}
{L{\"o}ptien}, B., {Gizon}, L., {Birch}, A.~C., {et~al.} 2018, Nature
  Astronomy, 2, 568

\bibitem[{{Martin-Belda} \&
  {Cameron}(2017{\natexlab{a}})}]{MartinBelda_2017_Evolution}
{Martin-Belda}, D. \& {Cameron}, R.~H. 2017{\natexlab{a}}, \aap, 603, A53

\bibitem[{{Martin-Belda} \&
  {Cameron}(2017{\natexlab{b}})}]{MartinBelda_2017_Inflows}
{Martin-Belda}, D. \& {Cameron}, R.~H. 2017{\natexlab{b}}, \aap, 597, A21

\bibitem[{{Nelson} {et~al.}(2013){Nelson}, {Brown}, {Brun}, {Miesch}, \&
  {Toomre}}]{Nelson_2013}
{Nelson}, N.~J., {Brown}, B.~P., {Brun}, A.~S., {Miesch}, M.~S., \& {Toomre},
  J. 2013, \apj, 762, 73

\bibitem[{{November} \& {Simon}(1988)}]{November_Simon_1988}
{November}, L.~J. \& {Simon}, G.~W. 1988, \apj, 333, 427

\bibitem[{Oliphant(2006)}]{oliphant2006guide}
Oliphant, T.~E. 2006, A guide to NumPy, Vol.~1 (Trelgol Publishing USA)

\bibitem[{{Roudier} {et~al.}(2013){Roudier}, {Rieutord}, {Prat}, {Malherbe},
  {Renon}, {Frank}, {{\v{S}}vanda}, {Berger}, {Burston}, \&
  {Gizon}}]{Roudier_2013}
{Roudier}, T., {Rieutord}, M., {Prat}, V., {et~al.} 2013, \aap, 552, A113

\bibitem[{{Schou} {et~al.}(2012){Schou}, {Scherrer}, {Bush}, {Wachter},
  {Couvidat}, {Rabello-Soares}, {Bogart}, {Hoeksema}, {Liu}, {Duvall}, {Akin},
  {Allard}, {Miles}, {Rairden}, {Shine}, {Tarbell}, {Title}, {Wolfson},
  {Elmore}, {Norton}, \& {Tomczyk}}]{Schou_2012}
{Schou}, J., {Scherrer}, P.~H., {Bush}, R.~I., {et~al.} 2012, \solphys, 275,
  229

\bibitem[{{Schunker} {et~al.}(2020){Schunker}, {Baumgartner}, {Birch},
  {Cameron}, {Braun}, \& {Gizon}}]{Schunker_2020}
{Schunker}, H., {Baumgartner}, C., {Birch}, A.~C., {et~al.} 2020, \aap, 640,
  A116

\bibitem[{{Schunker} {et~al.}(2019){Schunker}, {Birch}, {Cameron}, {Braun},
  {Gizon}, \& {Burston}}]{Schunker_2019}
{Schunker}, H., {Birch}, A.~C., {Cameron}, R.~H., {et~al.} 2019, \aap, 625, A53

\bibitem[{{Schunker} {et~al.}(2016){Schunker}, {Braun}, {Birch}, {Burston}, \&
  {Gizon}}]{Schunker_2016}
{Schunker}, H., {Braun}, D.~C., {Birch}, A.~C., {Burston}, R.~B., \& {Gizon},
  L. 2016, \aap, 595, A107

\bibitem[{{Sheeley}(1972)}]{Sheeley_1972}
{Sheeley}, N.~R., J. 1972, \solphys, 25, 98

\bibitem[{{Snodgrass}(1983)}]{Snodgrass_1983}
{Snodgrass}, H.~B. 1983, \apj, 270, 288

\bibitem[{{Spruit}(2003)}]{Spruit_2003}
{Spruit}, H.~C. 2003, \solphys, 213, 1

\bibitem[{Stix(2002)}]{Stix_2002}
Stix, M. 2002, The Sun, 2nd edn. (Springer), corrected second printing 2004

\bibitem[{{Thompson}(2006)}]{Thompson_2006}
{Thompson}, W.~T. 2006, \aap, 449, 791

\bibitem[{{van Driel-Gesztelyi} \& {Green}(2015)}]{Driel_Gesztelyi_2015}
{van Driel-Gesztelyi}, L. \& {Green}, L.~M. 2015, Living Reviews in Solar
  Physics, 12, 1

\bibitem[{{Vargas Dom{\'\i}nguez} {et~al.}(2008){Vargas Dom{\'\i}nguez},
  {Rouppe van der Voort}, {Bonet}, {Mart{\'\i}nez Pillet}, {Van Noort}, \&
  {Katsukawa}}]{Vargas-Dominguez_2008}
{Vargas Dom{\'\i}nguez}, S., {Rouppe van der Voort}, L., {Bonet}, J.~A.,
  {et~al.} 2008, \apj, 679, 900

\bibitem[{{Verma} {et~al.}(2013){Verma}, {Steffen}, \& {Denker}}]{Verma_2013}
{Verma}, M., {Steffen}, M., \& {Denker}, C. 2013, \aap, 555, A136

\bibitem[{Virtanen {et~al.}(2020)Virtanen, Gommers, Oliphant, Haberland, Reddy,
  Cournapeau, Burovski, Peterson, Weckesser, Bright, {van der Walt}, Brett,
  Wilson, Millman, Mayorov, Nelson, Jones, Kern, Larson, Carey, Polat, Feng,
  Moore, {VanderPlas}, Laxalde, Perktold, Cimrman, Henriksen, Quintero, Harris,
  Archibald, Ribeiro, Pedregosa, {van Mulbregt}, \& {SciPy 1.0
  Contributors}}]{2020SciPy-NMeth}
Virtanen, P., Gommers, R., Oliphant, T.~E., {et~al.} 2020, Nature Methods, 17,
  261

\bibitem[{{{\v{S}}vanda} {et~al.}(2013){{\v{S}}vanda}, {Roudier}, {Rieutord},
  {Burston}, \& {Gizon}}]{Svanda_2013}
{{\v{S}}vanda}, M., {Roudier}, T., {Rieutord}, M., {Burston}, R., \& {Gizon},
  L. 2013, \apj, 771, 32

\bibitem[{{{\v{S}}vanda} {et~al.}(2014){{\v{S}}vanda}, {Sobotka}, \&
  {B{\'a}rta}}]{Svanda_2014}
{{\v{S}}vanda}, M., {Sobotka}, M., \& {B{\'a}rta}, T. 2014, \apj, 790, 135

\bibitem[{{{\v{S}}vanda} {et~al.}(2007){{\v{S}}vanda}, {Zhao}, \&
  {Kosovichev}}]{Svanda_2007}
{{\v{S}}vanda}, M., {Zhao}, J., \& {Kosovichev}, A.~G. 2007, \solphys, 241, 27

\bibitem[{{Welsch} {et~al.}(2004){Welsch}, {Fisher}, {Abbett}, \&
  {Regnier}}]{Welsch_2004}
{Welsch}, B.~T., {Fisher}, G.~H., {Abbett}, W.~P., \& {Regnier}, S. 2004, \apj,
  610, 1148

\bibitem[{{W}es {M}c{K}inney(2010)}]{mckinney-proc-scipy-2010}
{W}es {M}c{K}inney. 2010, in {P}roceedings of the 9th {P}ython in {S}cience
  {C}onference, ed. {S}t\'efan van~der {W}alt \& {J}arrod {M}illman, 56 -- 61

\bibitem[{{Zhao} \& {Kosovichev}(2004)}]{Zhao_2004}
{Zhao}, J. \& {Kosovichev}, A.~G. 2004, \apj, 603, 776

\end{thebibliography}


\begin{appendix}

\section{Test of the effect of Zernike subtraction on the flow features}
\label{section_Appendix_synthetic_data}

\begin{figure}
\centering
\includegraphics[width=\hsize]{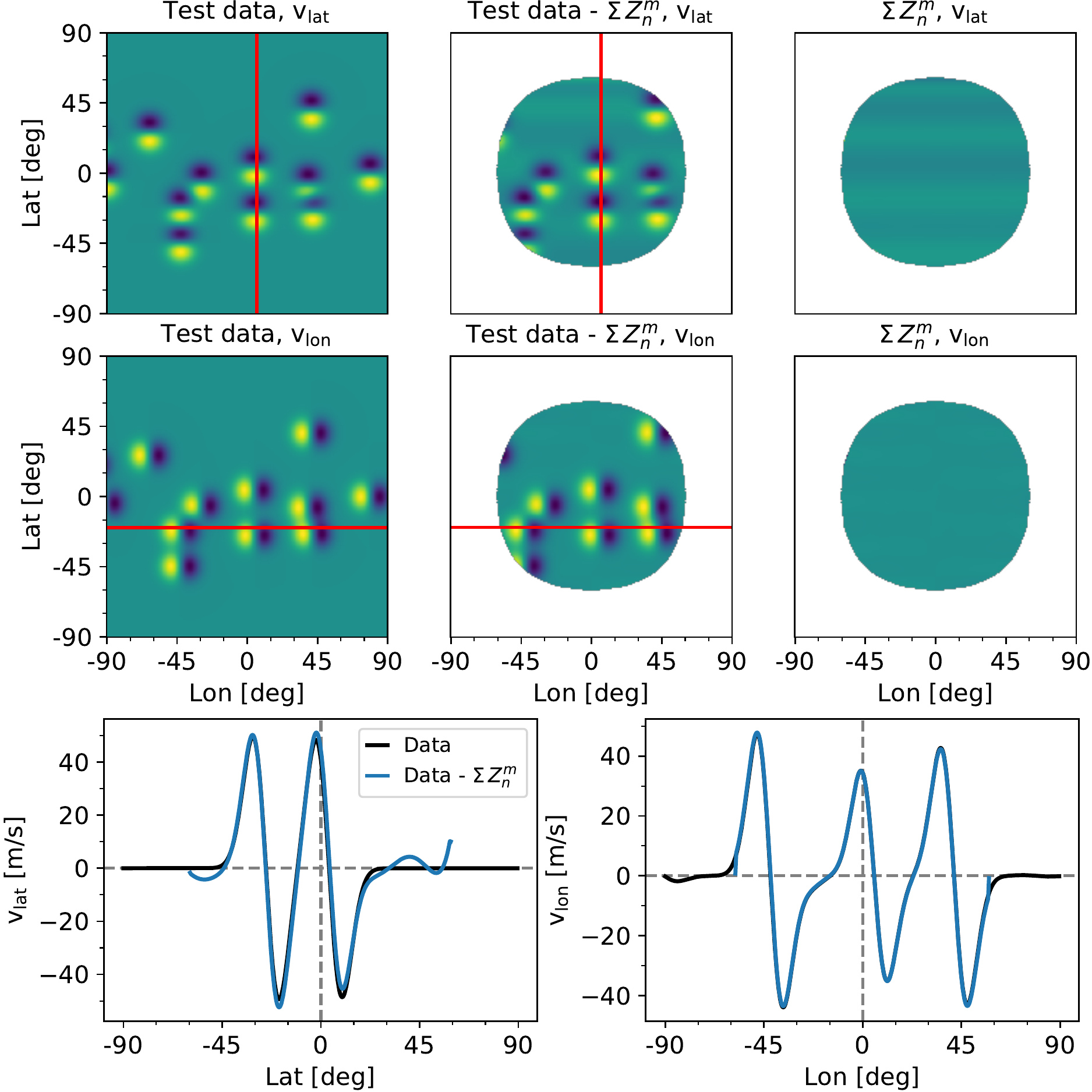}
  \caption{Example time step of the test with synthetic data. Top panel: Left column: Synthethic flow map in Plate Carree coordinates, for $v_{\rm lat}$ (top) and $v_{\rm lon}$ (bottom). Middle: After subtracting the Zernike fits $Z_n^m$ and limiting the field of view to \SI{60}{\degree} from disk center. Right: Sum over the Zernike fits. The red lines indicate the positions of the line plots in the bottom panel. Bottom panel: Line plots showing the synthetic data before (black) and after (blue) subtraction of the Zernike fits, for $v_{\rm lat}$(left) and $v_{\rm lon}$(right).}
      \label{Zernike_synthetic_15deg}
\end{figure}

We want to test the effect that the processing applied to the data (Sect.~\ref{subsec_processingvalidation}), specifically the subtraction of the fitted Zernike polynomials $Z_n^m$, has on flows resembling the inflows towards active regions that we investigate. For this, we construct synthetic flow maps, fit the Zernike polynomials to these maps, and subtract the Fourier-filtered components of the fitted Zernike time series. This is done in the same way as for the actual LCT data. We create one Carrington rotation of synthetic data, with frames set 30~minutes apart, by rotating the grid 0.272 degrees from one time step to the next. This corresponds to a rotation period of 27.5735~days.

To create the data, we impose synthetic flows on a Plate Carree coordinate grid (longitude, latitude) of size 450~x~450 with a grid spacing of \SI{0.4}{\degree}. The B~angle $B_0$ and the P~angle $\Phi_0$ are assumed to be zero, for the sake of simplicity. As a model for the flows, we use 2D~Gaussian derivatives in the longitudinal direction (for the longitudinal flow component) and in the latitudinal direction (for the latitudinal flow component). We create two different setups, a periodic case and a random case. In the periodic case, the flows are equally distributed on a grid along the equator and the central meridian. In the random case, the flows are distributed randomly within $\pm \SI{45}{\degree}$ latitude. The peak velocity of the flow features is \SI{50}{\meter \per \second} in all cases. We ran two different cases for both the periodic and the random setup, with widths of the flow features of approximately \SI{8} and \SI{12}{\degree} (which is where the velocities drop below \SI{20}{\meter \per \second}). In addition, we ran the same tests with additional moat flows imposed on the synthetic inflows, which are realized as Gaussians with widths of \SI{1.2}{\degree} and peak velocities of \SI{500}{\meter \per \second}. This is done to test whether the small-scale moat flow introduces additional deviations or leaking into the inflows.

First, we transform the velocities of the flow maps from m/s to pixel/s. This is done by inverting the equations given in the Appendix~C of \citet{Loeptien_2017}:
\begin{eqnarray}
v_n & = \dfrac{c_{11} v_{\theta} - c_{21} v_{\phi}} {c_{11} c_{22} - c_{12}c_{21}},\\
v_m & = \dfrac{c_{12} v_{\theta} - c_{22} v_{\phi}} {-c_{11} c_{22} + c_{12}c_{21}},
\end{eqnarray}\\
where we follow the definitions and nomenclature of \citet{Loeptien_2017}. We then map the Plate Carree coordinate grid (longitude, latitude) to a CCD coordinate grid ($x,y$) of size 1024~x~1024 grid points, which is the effective size that the processing pipeline uses for the HMI data. The projection is done using the transformation specified by \textit{sphere2img()} from the Stanford ring-diagram pipeline \citep{Bogart_2011a}.

We fit the Zernike polynomials, Fourier-filter them and subtract them from the data. Then, we re-map from the CCD back to the Plate Carree projection. This is done using the transformation specified by \textit{img2sphere()} from the Stanford ring-diagram pipeline \citep{Bogart_2011a}.

The last step is to retransform the velocities from pixel/s to m/s, by applying the equations from \citet{Loeptien_2017}.

Fig.~\ref{Zernike_synthetic_15deg} shows the first time step of the random test case with inflow extents of about \SI{12}{\degree}, before and after subtraction. The lower panels show cuts through the maps for $v_{\rm lon}$ and $v_{\rm lat}$.
Towards the observation limb at \SI{60}{\degree}, deviations from the synthetic data are noticeable, with velocities below \SI{5}{\meter \per \second}. Both the shape and extent of the flows are unaltered by the processing. The peak velocities are changed by less than \SI{10}{\percent}. The tests including moat flows showed no artifacts in addition to this.

\section{Comparison to flows from direct Doppler images}
\label{section_Appendix_vLOS_vDopp}

We compare the velocities obtained by LCT with those of direct Doppler images on an example active region. Comparisons of flows from helioseismic measurements to Doppler data were carried out by for example \citet{Gizon_2000}, \citet{Jackiewicz_2008} and \citet{Svanda_2013}. For the comparison, we transform the LCT flow maps from ($v_{\rm lon}$, $v_{\rm lat}$) to the line-of-sight (LOS) component. \citet{Roudier_2013} applied similar transformations to calculate the spherical components from $x$-, $y$- and Doppler components. We point out that the conventions used in our derivation below differ from the ones used by \citet{Roudier_2013}.

\subsection{Projection of LCT to the line of sight}
\label{Project_LCT_to_LOS}

We use a right-handed system in which $x$ points towards the observer and $z$ towards solar north. In this system, latitude $\lambda$ is defined between $[-90:90]$ and longitude $\varphi$ is defined between $[0:360)$. Disk center is thus at $\lambda=\SI{0}{\degree}, \varphi=\SI{0}{\degree}$ (when the B angle $B_0 = \SI{0}{\degree}$). The basis vectors of this system are:
\begin{eqnarray}
\vec{e_r} =
\begin{pmatrix}
 \cos(\lambda) \cos(\varphi)\\
\cos(\lambda) \sin(\varphi)\\
\sin(\lambda)
\end{pmatrix},
\vec{e_{\lambda}} =
\begin{pmatrix}
-\sin(\lambda) \cos(\varphi)\\
-\sin(\lambda) \sin(\varphi)\\
\cos(\lambda)
\end{pmatrix},
\vec{e_{\varphi}} =
\begin{pmatrix}
-\sin(\varphi)\\
\cos(\varphi)\\
0
\end{pmatrix}.
\end{eqnarray}
The transformation between Cartesian and spherical coordinates follows as
\begin{equation}
\begin{pmatrix}
\vec{e_x} \\ \vec{e_y} \\ \vec{e_z} 
\end{pmatrix} =
\begin{pmatrix}
\cos(\lambda) \cos(\varphi) & -\sin(\lambda) \cos(\varphi) & -\sin(\varphi)\\
\cos(\lambda) \sin(\varphi) & -\sin(\lambda) \sin(\varphi) & \cos(\varphi)\\
\sin(\lambda) & \cos(\lambda) & 0 
\end{pmatrix}
\begin{pmatrix}
\vec{e_r} \\ \vec{e_{\lambda}} \\ \vec{e_{\varphi}}
\end{pmatrix}.
\end{equation}
In addition, the rotation due to the changing B~angle $B_0$ has to be corrected for. This is a rotation around the $y$-axis. Thus
\begin{eqnarray}
\begin{pmatrix}
\vec{e_x} \\ \vec{e_y} \\ \vec{e_z} 
\end{pmatrix} =
\begin{pmatrix}
\cos \left(B_0 \right) & 0 & \sin \left(B_0 \right) \\
0 & 1 & 0 \\
-\sin \left(B_0 \right) & 0 & \cos \left(B_0 \right) \nonumber
\end{pmatrix} \times \\
\begin{pmatrix}
\cos(\lambda) \cos(\varphi) & -\sin(\lambda) \cos(\varphi) & -\sin(\varphi)\\
\cos(\lambda) \sin(\varphi) & -\sin(\lambda) \sin(\varphi) & \cos(\varphi)\\
\sin(\lambda) & \cos(\lambda) & 0 
\end{pmatrix}
\begin{pmatrix}
\vec{e_r} \\ \vec{e_{\lambda}} \\ \vec{e_{\varphi}}
\end{pmatrix}.
\end{eqnarray}
With LCT, we only have information on $\vec{e_{\lambda}} $ and $\vec{e_{\varphi}} $, and the radial component $ \vec{e_r} $ is assumed to be zero. Thus, projecting the LCT velocity vectors onto the line of sight $ \vec{e_x} $ follows the expression:
\begin{align}
\begin{split}
\vec{e_x} = & \left(-\sin \left(\lambda \right) \cos \left(\varphi \right) \cos \left(B_0 \right) + \cos \left(\lambda \right) \sin \left(B_0 \right) \right) \vec{e_{\lambda}} \\ &
+ \left(-\sin \left(\varphi \right)\cos \left(B_0 \right) \right)\vec{e_{\varphi}}
\end{split}.
\end{align}

\subsection{Data reduction of Doppler and line-of-sight data}

To compare the Doppler images to the line-of-sight projected LCT velocity images, large-scale systematics depending on disk position need to be removed. In addition, the data sets need to share a common frame of reference as well as temporal and spatial resolution.

First, we apply the same mapping procedure as was used for the magnetograms and the intensity maps to the full-disk Dopplergrams from hmi.v$\_$720s. This creates a Plate-Carree mapped cube for each EAR, with frames of \SI{60x60}{\degree} and a grid spacing of \SI{0.1}{\degree}. The cubes are centered on the coordinate specified in \citet{Schunker_2016}, and cover the time period between the start and end time provided in the HEAR survey. In this step, we also subtract the background signal that stems from solar rotation, which is taken from averages over a third of a Carrington rotation (hmi.v\_avg120). Each image is subtracted by the first average frame whose central time as specified by the \mbox{MidTime} keyword lies after the observation time as specified by the T\_REC keyword.

In addition to solar rotation, the observer motion in radial direction has to be corrected for. This is done by subtracting for each frame the velocity specified by the frame's OBS\_VR keyword.

Next, we average in time over one day, relative to the time of emergence, and apply Gaussian smoothing of $\sigma = \SI{0.4}{\degree}$. We then resample the grid spacing of the map from \SI{0.1}{\degree} to the \SI{0.4}{\degree} of the LCT velocity maps.

The convention for the sign of the HMI Dopplergrams is positive for movement away from the observer (redshift), and negative for movement towards the observer (blueshift). This is opposite to the convention used in the transformation in Sect.~\ref{Project_LCT_to_LOS}, where the $x$ axis points towards the observer. The sign of the Dopplergrams is therefore flipped for the further analysis.

\subsection{Results}

\begin{figure}
\centering
\includegraphics[width=\hsize]{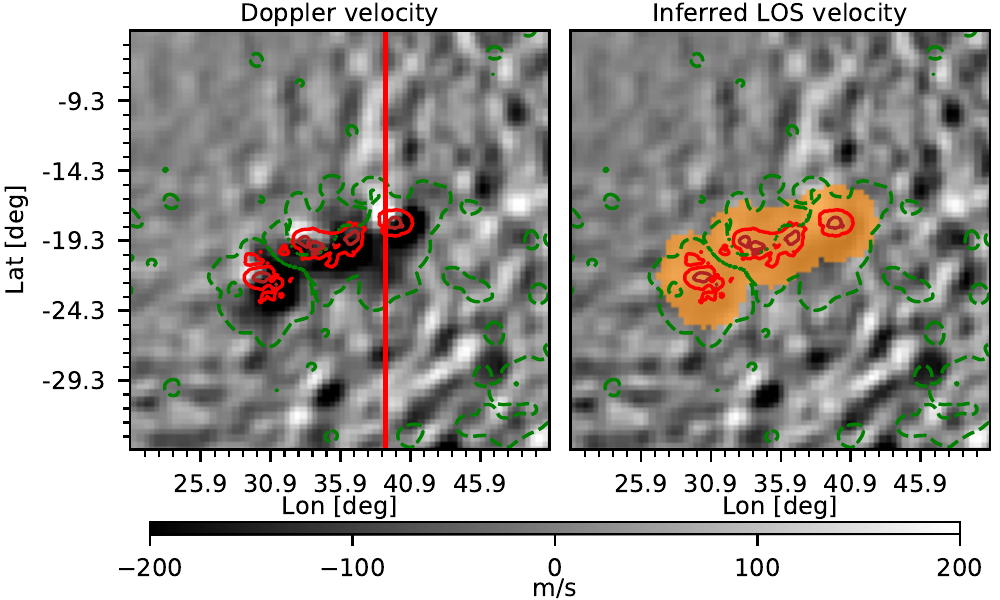}
  \caption{Comparison of Doppler velocity (left panel) and line-of-sight projected LCT velocity (right panel). The green contours outline the absolute radial magnetic field at \SI{10}{Gauss}.
  The dark- and bright-red contours indicate the umbral and penumbral boundary, respectively. The orange shaded region in the right panel indicates the region that is excluded in the following (see text for details). The red line in the left panel indicates the position of the line plot in
  Fig.~\ref{Comparison_vdopp_vlos_11158_lineplots}. The axis labels are in Carrington coordinates. The center of the maps is approximately \SI{23}{\degree} from disk center, which is in the direction towards the upper left corner.
}
      \label{Comparison_vdopp_vlos_11158}
\end{figure}

\begin{figure}
\centering
\includegraphics[width=\hsize]{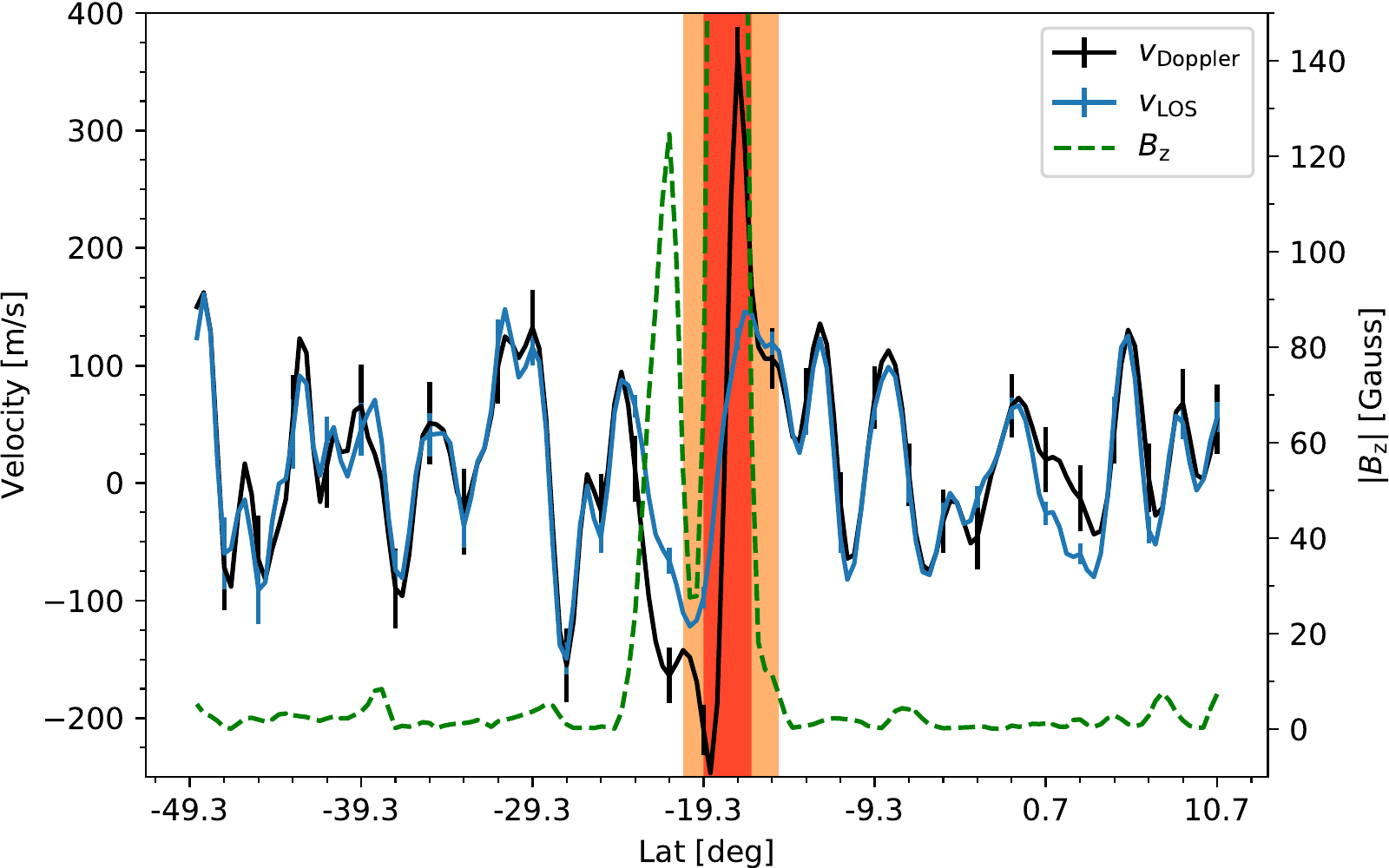}
  \caption{Plot along the red line indicated in Fig.~\ref{Comparison_vdopp_vlos_11158}. The black and blue lines indicate the velocity measurements from Doppler and line-of-sight, respectively. The errorbars indicate the standard error of the average. The dashed green line indicates the radial magnetic field strength. The red and orange shaded regions correspond to the red and orange regions in Fig.~\ref{Comparison_vdopp_vlos_11158}.
  }
      \label{Comparison_vdopp_vlos_11158_lineplots}
\end{figure}

\begin{figure}
\centering
\includegraphics[width=\hsize]{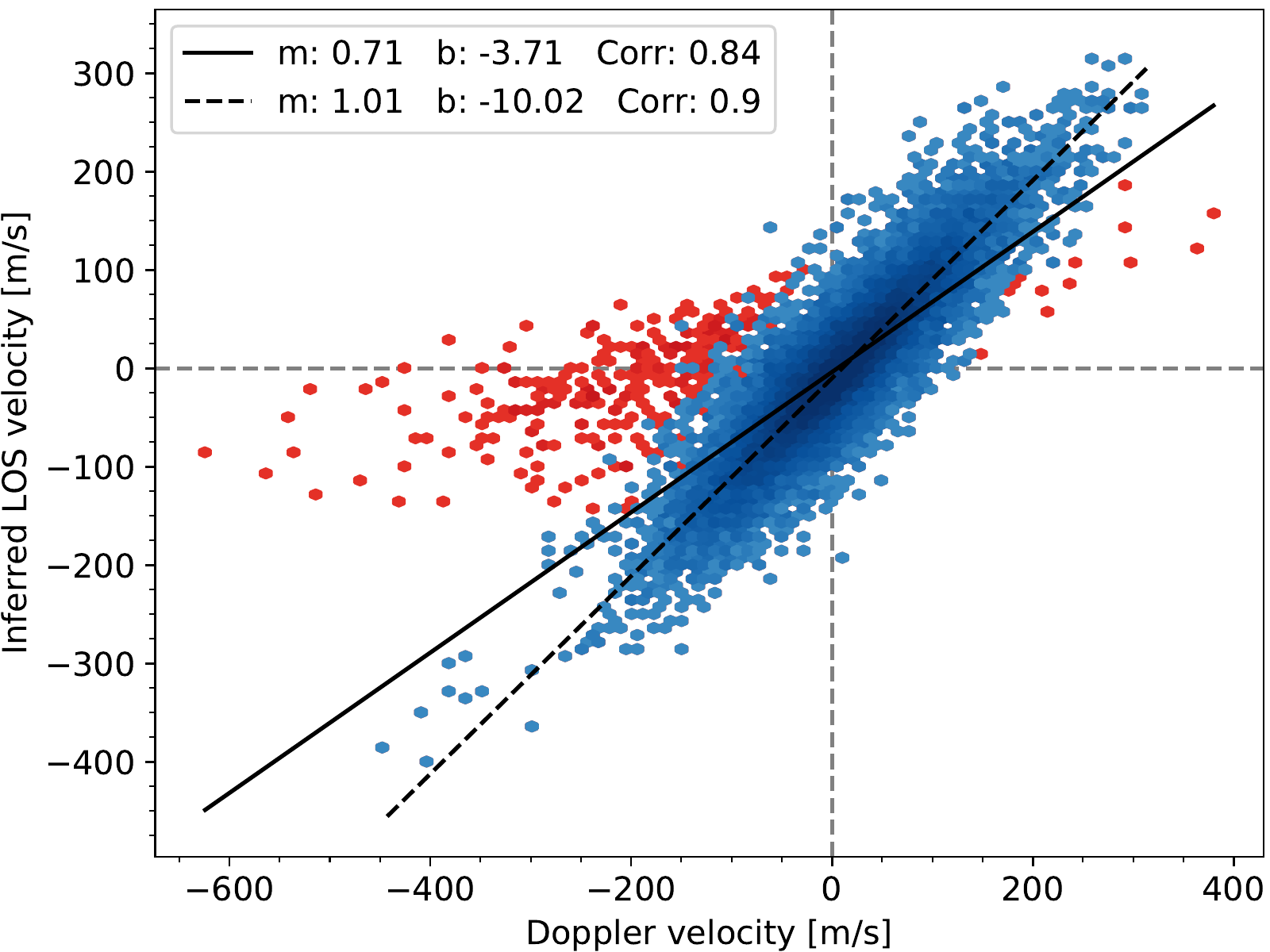}
  \caption{Density plot of the velocity data in Fig.~\ref{Comparison_vdopp_vlos_11158}. The red-shaded dots indicate all pixels, the blue-shaded dots only those outside of the orange shaded region in Fig.~\ref{Comparison_vdopp_vlos_11158}. The continuous (dashed) line indicates the fit to the full (magnetic-field-filtered) sample, respectively. The legend provides the fit parameters for both fits as well as the correlations between the measurements.}
      \label{Fit_vdopp_vlos_11158}
\end{figure}

Fig.~\ref{Comparison_vdopp_vlos_11158} shows the comparison of a Doppler map and the corresponding line-of-sight projected LCT map for AR~11158. The maps are averages over 24~hours, starting four days after the time of emergence.

Fig.~\ref{Comparison_vdopp_vlos_11158_lineplots} shows a cut along the red line in Fig.~\ref{Comparison_vdopp_vlos_11158}. Fig.~\ref{Fit_vdopp_vlos_11158} shows a density plot of the two velocities. We fit a line through the data for the relation between the Doppler and the LOS data. Because both are subject to error, we perform the fit with Orthogonal Distance Regression (ODR), with the standard error of the 24~hour time averages as errors. The errors close to disk center are on the order of \SI{15}{\meter \per \second} for the Doppler frames and \SI{5}{\meter \per \second} for the LOS frames, and increase outward to \SI{40}{\meter \per \second} and \SI{60}{\meter \per \second}, respectively. These are rough estimates, as the individual frames are not uncorrelated. We also compute the correlation between the two flow measurements.

The comparison (Figures \ref{Comparison_vdopp_vlos_11158},
\ref{Fit_vdopp_vlos_11158} and \ref{Comparison_vdopp_vlos_11158_lineplots}) illustrates that the Doppler velocity and the LOS velocity inferred from LCT agree well, except in the presence of strong magnetic field. Here, velocities from LCT are lower than the Doppler velocities. This has several reasons: LCT is known to underestimate velocities at strong magnetic field \citep{Fisher_Welsch_2008, Loeptien_2017}. Strong magnetic field suppresses convective blueshift, which results in a redshift-signature in the Doppler maps \citep{Stix_2002}. Furthermore, the two measurements are taken at different depths \citep{Jackiewicz_2008}.

When excluding regions of \SI{2}{\degree} around positions where the LCT flows are excluded due to reduced or increased continuum intensity (see Sect.~\ref{subsec_blankout}), the systematic outliers are ruled out and the slope of the fit between the Doppler- and the inferred line-of-sight velocities is closer to 1 (cf. Fig.~\ref{Fit_vdopp_vlos_11158}). The correlations between the Doppler- and the LOS velocities are between 0.7 and 0.9 for all observed one-day averages.

The intercept of the fit is in the range of \SIrange{-11}{-16}{\meter \per \second} for all one-day averages, meaning that the line-of-sight velocities are systematically lower than the Doppler velocities (i.e. the Doppler velocities have a preference towards upflows at disk center). Since the projection to the line of sight only takes into account $v_{\rm lon}$ and $v_{\rm lat}$, the line-of-sight velocity is necessarily zero at disk center, whereas the Doppler velocity is sensitive to any motion along the line of sight, including radial motions. Several studies investigated the rms velocity of Doppler images at disk center and the radial velocity of supergranulation, and reported velocities between \SI{5}{\meter \per \second} and \SI{15}{\meter \per \second} \citep{Giovanelli_1980, Hathaway_2002, Duvall_2010}. Our findings are in accordance with this.

\section{Comparison of methods to measure the positions of the AR}
\label{section_Appendix_Positions}

\begin{figure}
\centering
\includegraphics[width=\hsize]{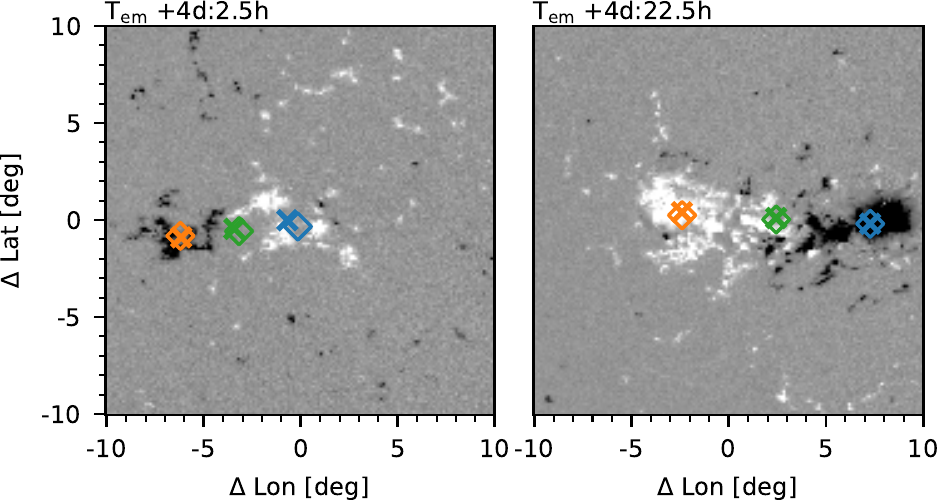}
\caption{Comparisons of the two position finding methods, on AR 11310 (left) and AR 11640 (right). The blue, orange, and green symbols mark the position of the leading and trailing polarity and the center of the AR, respectively. The crosses and diamonds refer to coordinates derived with the method described in Sect.~\ref{Sec_Positions} and the method described by \citet{Schunker_2019}, respectively. The axis labels are relative coordinates provided by the HEAR survey. The background images show the radial field $B_z$. The gray scale saturates at \SI{\pm500}{\Gauss}.} \label{find_center_of_gravity_comparison}
\end{figure}

We compare the two different methods which were used to measure the positions of the polarities of the active regions (cf. Sect.~\ref{Sec_Positions}). Fig.~\ref{find_center_of_gravity_comparison} shows two example active regions.

\section{Influence of moat flows}
\label{section_Appendix_SSQ}

\begin{figure}
\resizebox{\hsize}{!}
{\includegraphics[width=0.2\hsize]{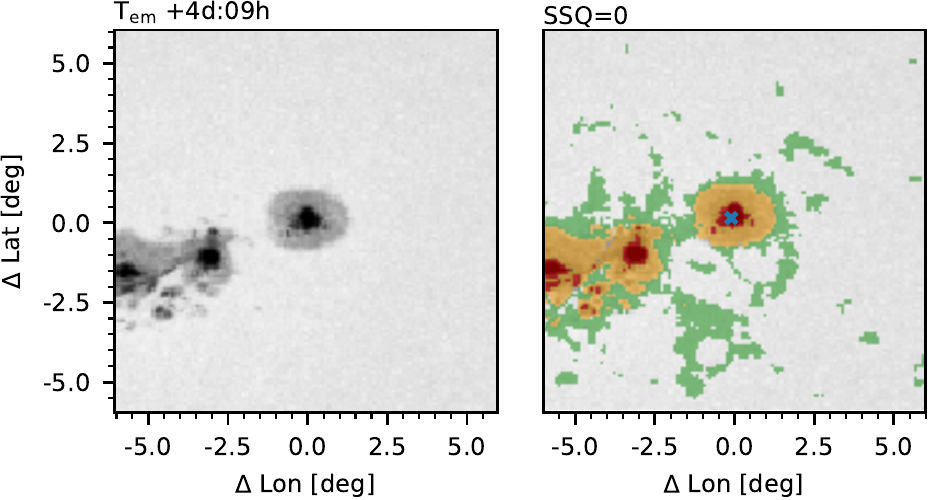}}
  \caption{Example of the sunspot classification. Left: A normalized six~hour average continuum map of AR~11158. The axis labels are relative to the center of the leading polarity, as defined in Sect.~\ref{Sec_Positions}. Right: Same as left, with colored overlay of pixels for which $|B_z|$ is higher than \SI{10}{\Gauss}. Colors indicate no intensity darkening (green), penumbra (yellow) and umbra (red) (see text for definition). The blue cross indicates the center of the largest umbral region. The time step shown here has a sunspot classification SSQ=0.
	  }
      \label{SSQ_exampleframe}
\end{figure}

\begin{figure*}
\centering
{\includegraphics[width=\hsize]{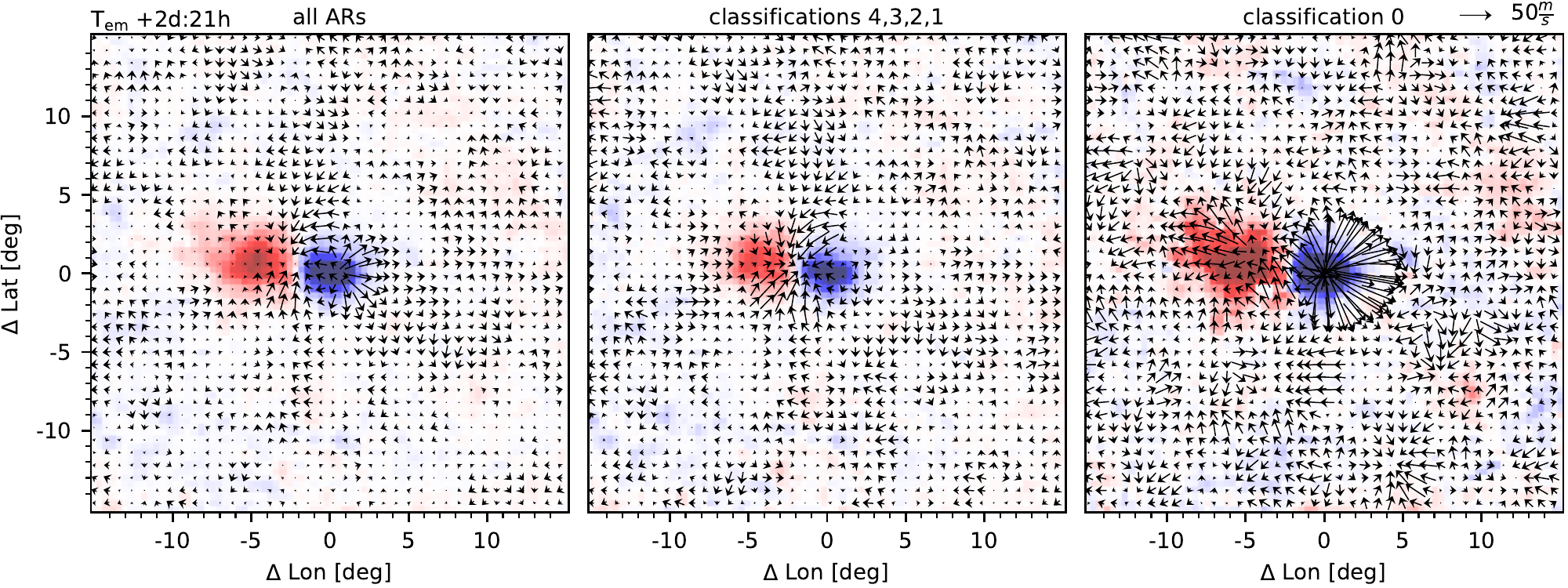}}
  \caption{Six~hour average flow maps, centered at 2~days and 21~hours after emergence. The left panel shows the average of 161 active regions, the middle panel shows the average over 119 active regions with no or only partial sunspot presence, and the right panel shows the average over 42 active regions with a sunspot with clear penumbra (corresponding to classification~0, see text). The active regions are averaged relative to the position of their leading polarity. Red (blue) indicates positive (negative) radial magnetic field saturated at $\pm$\SI{150}{\Gauss}. The arrows indicate the flows.
	  }
      \label{moat_flow_example}
\end{figure*}

We want to identify the presence of moat flows in the flow data on our sample of active regions. For this, we classify time series of six-hour averages of each active region in the sample according to the presence of a sunspot with clear penumbra, which would potentially host a moat flow. We then ensemble average the resulting subsamples to identify the moat flow signature.

The classification is done with the following procedure:
\begin{itemize}
\item Crop to the inner \SI{12x12}{\degree} of the normalized continuum intensity frames and the magnetograms around the position of the leading polarity.
\item Create a mask of all pixels with the absolute radial field $|B_z|$ below \SI{10}{\Gauss} and apply that mask to the continuum frame. This is used as the total number of pixels corresponding to the active region.
\item Count the number of pixels in the normalized continuum intensity frame that fall in bins of below 0.6 (corresponding to umbra), between 0.6 and 0.95 (penumbra), and above 0.95 (no intensity darkening). The threshold at 0.95 is the same as was used for the exclusion of pixels in Sect.~\ref{subsec_blankout}, the one at 0.6 is a result of testing.
\end{itemize}

Connected regions of pixels with normalized intensity below 0.6 are identified as umbrae. For the largest one, the total pixel count of umbra and the percentage of penumbra in a box of \SI{2.2}{\degree} around the center of the umbra are calculated. Each time step is then classified into one of five categories with the following scheme: 
\begin{itemize}
\item 4 \textit{(No intensity darkening)} if the total number of pixels in the umbra is zero and the total percentage of pixels in the penumbra, relative to the total number of active region pixels, is less than 1.
\item 3 \textit{(Small pore)} if the total number of pixels in the umbra is zero and the total percentage of pixels in the penumbra is larger than 1.
\item 2 \textit{(Large pore, no penumbra)} if the number of pixels in the umbra of the largest spot is less than twelve and the percentage of penumbra around this umbra is less than 20.
\item 1 \textit{(Large pore, with some penumbra)} if either the number of pixels in the umbra of the largest spot is less than twelve and the percentage of penumbra around this umbra is larger than 20, or if the number of pixels in the umbra of the largest spot is larger than twelve and the percentage of penumbra around this umbra is less than 20.
\item 0 \textit{(Clear penumbra)} if the number of pixels in the umbra of the largest spot is larger than twelve and the percentage of penumbra around this umbra is larger than 20.
\end{itemize}

This classification scheme was derived and checked by inspecting several ARs. A table of all time steps between -2 and +7~days for all 182 active regions is available online.

Fig.~\ref{SSQ_exampleframe} shows an example of the classification. The colored regions in the right panel indicate pixels with an absolute radial field above \SI{10}{\Gauss}, which we in this context associate with the AR. The small features at large distances from the sunspots do not contribute significantly.

Fig.~\ref{moat_flow_example} shows an example time step for averages over three different sets of data: All 161~active regions for which data is available at this time relative to emergence, only those regions without a sunspot with clear penumbra (classifications of~1,2,3~and~4), and only those regions with a sunspot with clear penumbra (classification of~0). It is apparent that the diverging flow at the leading polarity is connected to the presence of developed sunspots with a clear penumbra, which is consistent with the moat flow. The velocities are around \SI{150}{\meter \per \second}, which is lower than typical moat flow velocities observed in evolved sunspots. A reason for this could be that the sunspots in our sample are just forming, together with the (in this context) comparatively large grid spacing of \SI{0.4}{\degree} and additional Gaussian smoothing of $\sigma = \SI{0.8}{\degree}$ of our flow data, which reduces localized peak velocities.

Also, the moat flow seen here is not symmetric: The westward (i.e. prograde) component is stronger than the eastward (retrograde) component. The components in the north- and southward directions have approximately the same strength as the westward component. \citet{Svanda_2014} also found an asymmetry between the east- and westward moat flows.

Our findings are in agreement with \citet{Vargas-Dominguez_2008}, who showed that moat flows are predominantly present in the directions where there is penumbra. The sunspots in our sample are only just forming, with additional flux joining the sunspot from the emergence site, that is, the retrograde direction. Once the accumulation of flux is ended and the penumbra surrounds the umbra in all directions, the moat flows become more symmetric. At later times however, the number of ARs that host a spot decreases, leading to high background noise.

\section{Average over all EARs}
\label{section_Appendix_fullensemble}

\subsection{Flows during the time of emergence}

\begin{figure*}
  {\includegraphics[width=\textwidth]{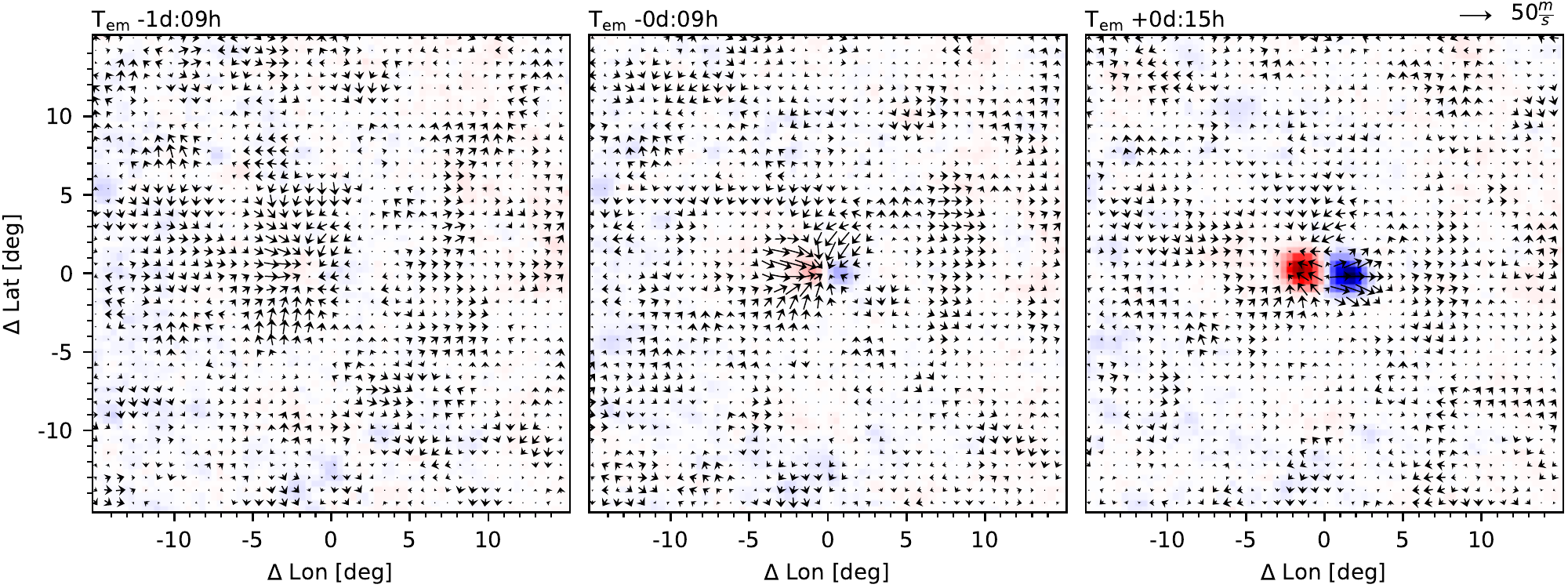}}
  \caption{Evolution (from left to right) during the emergence phase of the average active region, for the ensemble average over all 182~ARs in the sample. Each time step is an average over six~hours, centered on the labeled time. The flows are smoothed with a Gaussian of $\sigma = \SI{0.8}{\degree}$. Red (blue) indicates positive (negative) radial magnetic field, saturated at $\pm$\SI{300}{\Gauss}.
  }
     \label{Flows_at_timesteps_quiver}
\end{figure*}

We perform an ensemble average over all 182~EARs in the sample, to connect to the work of previous studies. In this case, we average in time over six~hours and apply Gaussian smoothing of $\sigma = \SI{0.8}{\degree}$ to the flow maps. Fig.~\ref{Flows_at_timesteps_quiver} shows the early evolution of flows in several time steps. Starting one day before emergence, a converging flow towards the center of the average EAR is clearly visible, becoming stronger towards the time of emergence. This converging flow is noticeable in all four flux-sorted subsamples (see Sect.~\ref{subsection_subsample_USFLUX}), and was found by \citet{Birch_2019}. During emergence, these flows vanish. At the same time, a prograde flow at the position of the leading polarity forms. This is again consistent with \citet{Birch_2019}.

\subsection{Flow variation in time, per polarity}

\begin{figure}
    {\includegraphics[width=\hsize]{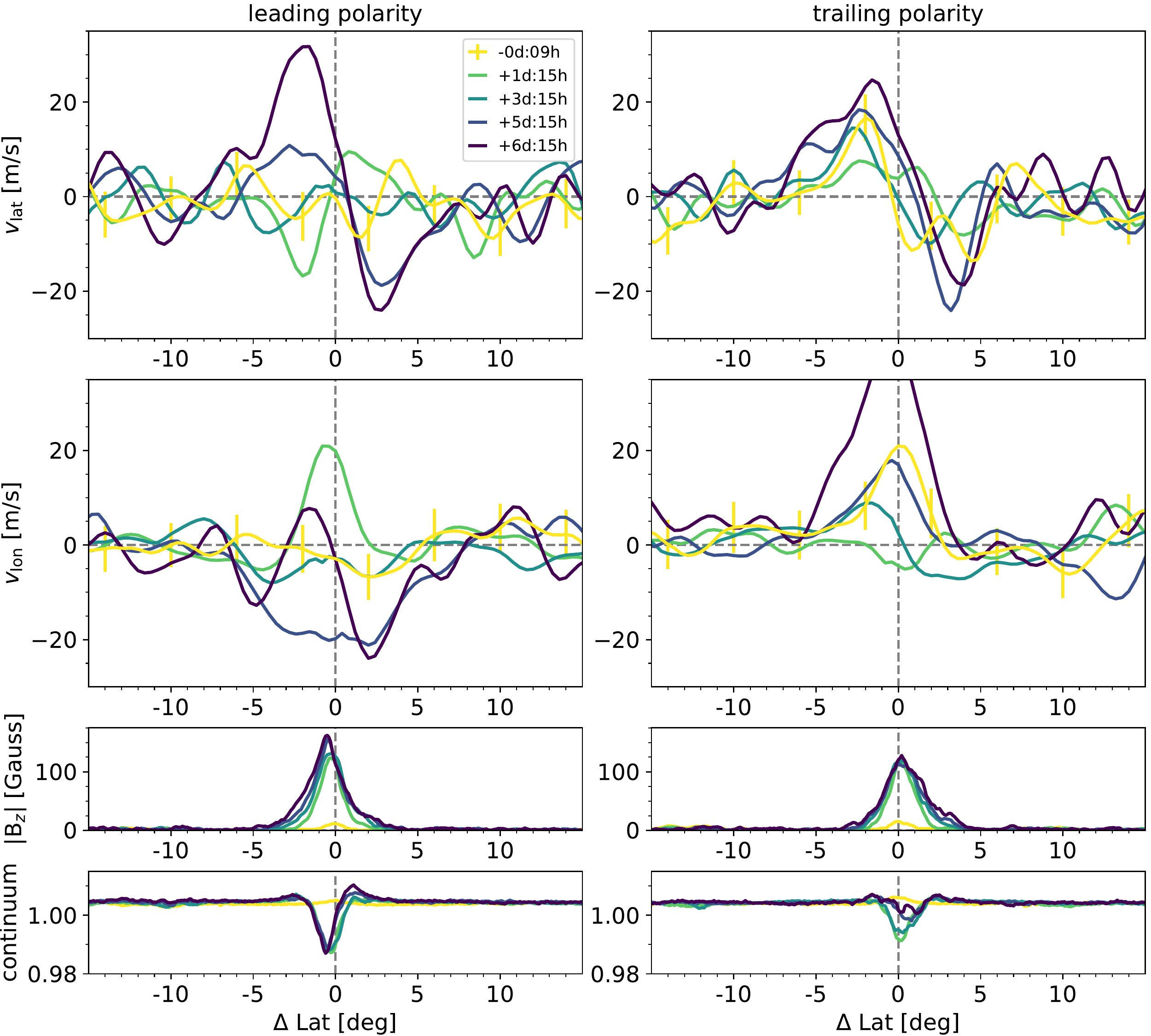}}
    \caption{Evolution of the flow in a longitudinal band of \SI{6}{\degree} width for each polarity. The left (right) column shows the leading (trailing) polarity. The first (second) row shows the latitudinal (longitudinal) flow component, the third row shows the absolute value of the radial magnetic field. The fourth row shows the normalized continuum intensity. All data are time averaged over six~hours. The flows are smoothed with a Gaussian of $\sigma = \SI{0.8}{\degree}$. The curves represent different time steps (see legend).
    }
 \label{timevariation_flows_blanked}
\end{figure}

Following the method of \citet{Braun_2019}, we calculate longitudinal averages of the flows. For each polarity, we average over \SI{6}{\degree} from the center of the active region towards east (west) for the trailing (leading) polarity.

Fig.~\ref{timevariation_flows_blanked} shows the results for the flows outside of strong magnetic field. The longitudinal flow components corresponding to time steps from three~days onwards show the retrograde flow that \citet{Braun_2019} detected. For the leading polarity, they span out to a maximum of approximately \SI{7}{\degree}, with velocities of about \SI{20}{\meter \per \second}. The velocities here are smaller than for the flux-binned subsamples in Sec.~\ref{subsection_subsample_USFLUX} because of the different averaging ranges.

\end{appendix}

\end{document}